\documentclass[trackchanges,twocolumn]{aastex701}

\usepackage{booktabs}
\usepackage{fancyhdr}
\usepackage{threeparttable}
\usepackage{siunitx}
\usepackage{amsmath,amsfonts,amssymb}
\usepackage{mathtools}
\usepackage{physics}
\usepackage{bm}
\usepackage{enumerate}

\begin{document}

\title{Neural-Network Chemical Emulator for First-Star Formation: Robust Iterative Predictions over a Wide Density Range}

\author[orcid=0009-0007-8748-2717,gname=Sojun,sname='Ono']{Sojun Ono}
\affiliation{Department of Astronomy, Kyoto University, Kitashirakawa-Oiwake-cho, Sakyo-ku, Kyoto 606-8502, Japan}
\affiliation{Faculty of Science, Hokkaido University, Sapporo, Hokkaido 060-0810, Japan}
\email[show]{ono@kusastro.kyoto-u.ac.jp}  

\author[orcid=0000-0001-7842-5488,gname=Kazuyuki, sname='Sugimura']{Kazuyuki Sugimura} 
\affiliation{Faculty of Science, Hokkaido University, Sapporo, Hokkaido 060-0810, Japan}
\email[show]{sugimura@sci.hokudai.ac.jp}

\begin{abstract}
We present a neural-network emulator for the thermal and chemical evolution in Population~III star formation. The emulator accurately reproduces the thermochemical evolution over a wide density range spanning 21 orders of magnitude ($10^{-3}$--$10^{18}\,\mathrm{cm}^{-3}$), tracking six primordial species: H, H$_2$, e$^{-}$, H$^{+}$, H$^{-}$, and H$_2^{+}$. To handle the broad dynamic range, we partition the density range into five subregions and train separate deep operator networks (DeepONets) in each region. When applied to randomly sampled thermochemical states, the emulator achieves relative errors below $10\,\%$ in over $90\,\%$ of cases for both temperature and chemical abundances (except for the rare species H$_2^{+}$). The emulator is roughly ten times faster on a CPU and more than 1000 times faster for batched predictions on a GPU, compared with conventional numerical integration. Furthermore, to ensure robust predictions under many iterations, we introduce a novel timescale-based update method, where a short-timestep update of each variable is computed by rescaling the predicted change over a longer timestep equal to its characteristic variation timescale. In one-zone collapse calculations, the results from the timescale-based method agree well with traditional numerical integration even with many iterations at a timestep as short as $10^{-4}$ of the free-fall time. This proof-of-concept study suggests the potential for neural network-based chemical emulators to accelerate hydrodynamic simulations of star formation.
\end{abstract}


%%%%%%%%%%%%%%%%%%%%%%%%%%%%%%%%%%%%%%%%%%%%%%%%%%%%%%%%%%%%%%%%%%%%%%%%%%
%%%%%%%%%%%%%%%%%%%%%%%%%%%%%%%%%%%%%%%%%%%%%%%%%%%%%%%%%%%%%%%%%%%%%%%%%%

\section{Introduction}
\label{sec:introduction}
In the Universe, gas collapses under gravity to form stars, which then disperse their material back into the interstellar medium via stellar winds and supernovae explosions. Over cosmic time-scales, this cycle has been repeated, gradually enriching the gas with metals produced in stellar interiors. Therefore, elucidating the physical properties and formation mechanisms of the first-generation stars, known as Population III (Pop III) stars, which emerged from primordial, metal-free gas at the onset of this cosmic cycle, is a fundamental challenge in modern astrophysics \citep[for a recent review, see][]{Klessen:2023}. To gain a comprehensive understanding of Pop III star formation, it is essential to combine hydrodynamical simulations with detailed thermochemical evolution calculations \citep[e.g.,][]{Abel:2002, Bromm:2002, yoshidaProtostar2008, hosokawaProtostellar2011, Stacy:2012, Susa:2013, Sugimura:2020}.

However, modeling the thermochemical evolution involved in the formation of Pop III stars is computationally challenging. These calculations require solving stiff differential equations, which are often handled using implicit methods involving iterative solvers \citep[e.g.,][]{omukaiPrimordial2001}. As the stiffness increases, especially in high-density environments, the number of iterations needed for implicit updates can grow substantially, leading to a high computational cost. Furthermore, the number of iterations differs across cells, resulting in load imbalances that reduce parallel efficiency in large-scale simulations. When accounting for metal-enriched environments with more complex chemical networks, the computational difficulty becomes even more severe.

In recent years, machine learning (ML) using neural networks (NNs) has demonstrated its usefulness in wide scientific applications, including the solution of differential equations \citep{navarroSolving2023}. ML has the potential to reduce computational costs, as it can infer solutions quickly and with fixed computational time once a model has been trained. Therefore, it is promising to construct emulators that accurately reproduce thermochemical evolution while avoiding the computational bottlenecks associated with implicit solvers. 

A growing number of studies have been exploring the application of ML to emulate thermochemical evolution. After the seminal work in this direction by \citet{Grassi:2011}, dimensionality reduction techniques using autoencoders have been widely studied \citep{holdshipChemulator2021,grassiReducing2022,sulzerSpeeding2023,maesMACE2024}. In related fields, ML has also been studied to improve the interpretation of ISM observations \citep{deMijolla:2019,Heyl:2023,Palud:2023} and to model networks of nuclear reactions inside stars \citep{grichenerNuclear2025}.

Yet, the above autoencoder-based ML methods have been applied to limited situations: \citet{holdshipChemulator2021} developed an emulator for a fixed time interval; \citet{grassiReducing2022} and \citet{sulzerSpeeding2023} constructed models for fixed density and temperature; and the model by \citet{maesMACE2024} is specialized for a one-dimensional model of AGB outflows. None of these approaches is suitable for modeling chemical evolution during star formation, where physical states and timescales vary widely.

Recently, L.~Branca and A.~Pallottini employed a physics-informed neural network (PINN) \citep{raissiPhysicsinformed2019} in \citet{brancaNeural2023}, and later adopted deep operator network (DeepONet) \citep{luLearning2021} in \citet{brancaEmulating2024} to model the evolution of the interstellar medium (ISM) in a more general setup. The latter study achieved higher accuracy than the former, reporting both accurate predictions and a speed-up of up to 128 times compared to traditional methods. 

However, \citet{borBridging2025} found that the method presented in \citet{brancaEmulating2024} is prone to instability due to the accumulation of prediction errors. The problem of iterative use of emulators has been recognized and partially addressed in the context of autoencoder-based methods. This issue has been mitigated either by injecting Gaussian noise during training \citep{holdshipChemulator2021} or by training the iterative updates themselves \citep{maesMACE2024}. Moreover, no solution has yet been proposed for the method of \citet{brancaEmulating2024}, even though numerous iterations are inevitable when applying ML methods to hydrodynamic simulations.

Moreover, before applying the method of \citet{brancaEmulating2024} to the case of star formation, another major challenge remains. While their model is not constrained by the aforementioned limitation of previous autoencoder-based ML methods \citep[e.g.,][]{holdshipChemulator2021,grassiReducing2022,sulzerSpeeding2023,maesMACE2024}, it still covers only low-density regimes typical of the ISM. To fully capture the star formation process, emulators must operate over a much wider density range, extending into the high-density protostellar regime.

The objective of this study is to take a first step toward fast and accurate emulation of thermochemical evolution in the star formation process using NNs. To this end, we extend the method of \citet{brancaEmulating2024} in two directions.

First, we develop NN models that emulate the thermochemical evolution of gas during Pop III star formation, ranging from the low-density regime at the onset of gravitational collapse to the high-density regime associated with protostar formation. To address the difficulty of training NNs over the extremely wide density range encountered in star formation, we divide the full range into multiple subregions and train a separate NN in each subregion.

Second, we develop a new update method that ensures robust evolution under iteration of short-timestep predictions. In this method, each variable is updated by rescaling its predicted change over a characteristic timescale of significant variation if the timestep is so short that the physical change over the step is smaller than the typical prediction error. The characteristic timescale is estimated by auxiliary NNs.

In this proof-of-concept study, we aim to demonstrate that ML methods can be applied to model the thermochemical evolution during star formation. Therefore, we focus on the relatively simple case of metal-free, radiation-free primordial gas, and demonstrate the feasibility of using ML to emulate thermochemical evolution under these conditions. In the future, we also intend to extend this framework to more complex chemical networks that include metals and radiation.

This paper is organized as follows.  In Section~\ref{methods}, we describe the basic equations for the thermochemical evolution and our NN methods to emulate the evolution. In Section~\ref{results}, we present the results of emulation. Section~\ref{discussion} evaluates the acceleration achieved by NNs and compares our results with previous studies. Finally, in Section~\ref{sec:conclusion}, we summarize our findings and outline future directions.

Throughout this paper, we use $\log$ to denote $\log_{10}$ for brevity.

\begin{figure*}[htb!]
 \begin{center}
   \includegraphics[width=10 cm]{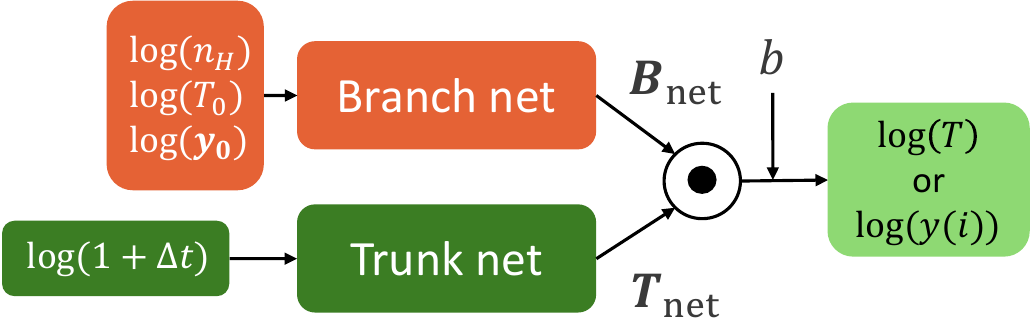}
 \end{center}
 \caption{Schematic diagram of DeepONet, which consists of two networks: the branch net and the trunk net. 
The branch net receives the hydrogen nucleus number density $n_\mathrm{H}$, the initial temperature $T_{0}$, and the initial chemical abundances $\bm{y}_{0}$ as input, in the form of $\log (n_\mathrm{H}/\mathrm{cm}^{-3})$, $\log (T_{0}/\mathrm{K})$, and $\log (\bm{y}_{0})$, respectively. The trunk net takes the timestep $\Delta t$ as input in the form of $\log\bigl(1 + \Delta t/\mathrm{sec}\bigr)$. The final output is computed as the inner product of the branch net output $\bm{B}_\mathrm{net}$ and the trunk net output $\bm{T}_\mathrm{net}$, plus a bias term $b$: $\bm{B}_\mathrm{net}\cdot\bm{T}_\mathrm{net}+b$. A separate DeepONet model is constructed for each target variable, which is either the updated temperature $T$ in the form of $\log (T/\mathrm{K})$ or the updated chemical abundance of the $i$-th species $y(i)$ in the form of $\log (y(i))$.
 }
 \label{fig:DeepONet}
\end{figure*}

%%%%%%%%%%%%%%%%%%%%%%%%%%%%%%%%%%%%%%%%%%%%%%%%%%%%%%%%%%%%%%%%%%%%%%%%%%
%%%%%%%%%%%%%%%%%%%%%%%%%%%%%%%%%%%%%%%%%%%%%%%%%%%%%%%%%%%%%%%%%%%%%%%%%%

\section{Methods}\label{methods}

\subsection{Basic Equations}
\label{sec:basic-equations}
We investigate the thermochemical evolution of primordial gas clouds during Pop III star formation. To model this evolution, we solve a set of ordinary differential equations (ODEs) governing the time evolution of the gas temperature and the abundances of key chemical species.

Our chemical model is based on \citet{Sugimura:2020, Sugimura:2023}, with additional high-density reactions \citep{Sadanari:2021}. The model tracks six chemical species: H, H$_2$, e$^-$, H$^+$, H$^-$, and H$_2^+$, which are involved in a network of 18 chemical reactions (see Appendix~\ref{sec:app-chemical-reactions} for details).  We focus on the pre-protostellar collapse phase and assume no external radiation field.

Heating and cooling processes are modeled as follows. For line cooling, we consider H$_2$ line cooling and Ly$\alpha$ cooling. We compute the H$_2$ line cooling rate using the optically thin cooling function from \citet{Glover:2015} with an average line escape probability obtained in \citet{Fukushima:2018}. For continuum cooling, we consider H free–bound emission, H$^-$ free–bound and free–free emission, H free–free emission, H$_2$–H$_2$ collision-induced emission (CIE), and H$_2$–He CIE. We calculate the cooling rate using the optical depth estimated with the Jeans length \citep[see][for details]{Tanaka:2014}, with continuum opacity from \citet{Matsukoba:2019}. Chemical heating and cooling are also included.

To simplify the model, we track only hydrogen-bearing species. Although deuterium chemistry, particularly HD cooling, plays a crucial role in the formation of low-mass Pop III stars \citep{Nagakura:2005}, we focus on the typical high-mass case and omit deuterium. We also neglect lithium chemistry, which does not affect thermal evolution but can alter the ionization fraction and, consequently, non-ideal magnetohydrodynamic (MHD) effects in magnetized environments \citep{Nakauchi:2019,Sadanari:2023}. Helium is assumed to remain fully neutral, with a fixed abundance of $y_\mathrm{He} = n_\mathrm{He}/n_\mathrm{H} = 9.7 \times 10^{-2}$, where $n_\mathrm{H}$ and $n_\mathrm{He}$ are the number densities of hydrogen and helium nuclei, respectively.

The basic equations for the time evolution of the chemical abundances and gas temperature are given by
\begin{align}
\dot{y}(i)&=\sum_jA_{ij}(n_\mathrm{H},T,\bm{y})\,, \label{ydot}\\
n_\mathrm{H}c_v \dot{T} &= \Gamma(n_\mathrm{H},T,\bm{y}) - \Lambda(n_\mathrm{H},T,\bm{y})\,, \label{Tdot}
\end{align}
where $\bm{y}$ is the vector of chemical abundances, and each abundance $y(i)$ is defined as $y(i)=n(i)/n_\mathrm{H}$, with $n(i)$ being the number density of species $i$. The index $i$ runs from 1 to 6 and corresponds to H, H$_2$, e$^-$, H$^+$, H$^-$, and H$_2^+$, respectively. Note that $n(1)=n(\mathrm{H})$ denotes the number density of neutral hydrogen, which is different from $n_\mathrm{H}$, the total number density of hydrogen nuclei. $A_{ij}$ denotes the contribution of reaction $j$ to $\dot{y}(i)$. The gas temperature is denoted by $T$, and $c_v$ is the specific heat at constant volume per hydrogen nucleus, depending on both $T$ and $\bm{y}$ \citep{omukaiFormation1998}. The functions $\Gamma(n_\mathrm{H},T,\bm{y})$ and $\Lambda(n_\mathrm{H},T,\bm{y})$ represent the heating and cooling rates, respectively.  Compressional heating is excluded from $\Gamma$, as it is treated separately by operator splitting when gas dynamics is included.

In this work, we mainly focus on the thermochemical evolution at densities below $n_\mathrm{H}=10^{18}\,\mathrm{cm^{-3}}$, where non-equilibrium chemistry and radiative (line and continuum) cooling are essential \citep{omukaiPrimordial2001}. At higher densities, we do not need to solve Eqs.~\eqref{ydot} and \eqref{Tdot} explicitly because reaction timescales become so short that chemical equilibrium is a good approximation, and the gas becomes so optically thick that radiative cooling is negligible and the gas can be treated as adiabatic. In this regime, the thermal and chemical state can be obtained from the Saha equation \citep{sahaLIII1920} with a given total energy. Accordingly, for $n_\mathrm{H}>10^{18}\,\mathrm{cm^{-3}}$, instead of constructing a chemical emulator, we model the thermochemical evolution by interpolating a precalculated table of temperature and equilibrium abundances as a function of density and total energy (see Appendix \ref{sec:app-table-equilibrium} for details). This method can significantly reduce the computational cost compared with conventional numerical integration.

We train the NN emulator using a set of physical states that broadly cover the thermal and chemical trajectories during Pop III star formation, which are obtained from one-zone calculations of a gravitationally collapsing gas cloud. The same trajectories are used to validate the emulator after training. The one-zone model follows the evolution of the central core, assuming free-fall collapse, $\dot{n}_\mathrm{H} = n_\mathrm{H} / t_\mathrm{ff}$, where the free-fall time is given by $t_\mathrm{ff} = (3\pi / (32 G \rho))^{1/2}$, with the mass density $\rho = (1 + 4 y_\mathrm{He})\, n_\mathrm{H} m_\mathrm{H}$ and the hydrogen mass $m_\mathrm{H}$.

The calculation starts from conditions corresponding to Pop III star formation at $z=16$ \citep{sugimuraCritical2014}: $n_\mathrm{H}=4.5\times10^{-3}\,\mathrm{cm^{-3}}$, $T=21\,\mathrm{K}$, $y(\mathrm{e}^-)=3.7\times10^{-4}$, and $y(\mathrm{H}_2)=2\times10^{-6}$. We evolve the system until $n_\mathrm{H}=10^{18}\,\mathrm{cm^{-3}}$, below which the chemical evolution must be treated in non-equilibrium. At each timestep, we (i) update $T$ and $\bm{y}$, (ii) add compressional heating $\Gamma_\mathrm{comp}=(\sum_iy(i) + y_\mathrm{He})k_\mathrm{B}\,\dot{n}_\mathrm{H}\,T$, where $k_\mathrm{B}$ is the Boltzmann constant, and (iii) increase the density.

In this work, we test four timestep sizes: $\Delta t = 10^{-1}$, $10^{-2}$, $10^{-3}$, and $10^{-4}\,t_\mathrm{ff}$. In a one-zone model, the relevant timescales are the free-fall time $t_\mathrm{ff}$ and the thermal and chemical timescales. However, in multi-zone hydrodynamic simulations, the timestep can be much shorter than these timescales, for instance, when the Courant condition imposed by small adaptive-mesh-refinement cells limits the timestep for the entire computational domain \citep[e.g.,][]{Sugimura:2020,Sugimura:2023}. In cases with small timesteps, such as $\Delta t \ll t_\mathrm{ff}$, the physical change per step can be comparable to or even smaller than the typical NN prediction error. Consequently, iterative NN updates can be dominated by the error and result in unphysical predictions \citep{borBridging2025}. We address this problem by introducing a new method in Section~\ref{section:timescale-based_method}.

%%%%%%%%%%%%%%%%%%%%%%%%%%%%%%%%%%%%%%%%%%%%%%%%%%%%%%%%%%%%%%%%%%%%%%%%%%
%%%%%%%%%%%%%%%%%%%%%%%%%%%%%%%%%%%%%%%%%%%%%%%%%%%%%%%%%%%%%%%%%%%%%%%%%%

\subsection{Neural Network for Emulating Thermal and Chemical Evolution: DeepONet}
\label{sec:deeponet}
We emulate the thermochemical evolution given by Eqs.\eqref{ydot} and \eqref{Tdot} employing DeepONet. DeepONet is a NN architecture originally developed to learn nonlinear operators and has recently been applied to solving differential equations. In astrophysics, \citet{brancaEmulating2024} successfully used DeepONet to model the thermochemical evolution of the ISM. We follow their methodology and adapt the network to the context of Pop III star formation. Our implementation uses the DeepXDE library \citep{luDeepXDE2021} built on the TensorFlow framework \citep{tensorflow2015-whitepaper}, which includes various DeepONet-related modules.

The overall structure of DeepONet is illustrated in Figure~\ref{fig:DeepONet}. A key feature of DeepONet is that it consists of two separate subnetworks: \textit{the branch net} and \textit{the trunk net}. Each is a standard feedforward NN with fully-connected layers, where each layer performs the transformation:
\begin{align}
    \bm{\phi}_{l} = \sigma\left(\bm{\phi}_{l-1}\bm{W}_l + \bm{b}_l\right)~~(l=1,2,\dots,N_l),
    \label{eq:one_layer}
\end{align}
where $\bm{\phi}_l \in \mathbb{R}^{d_l}$, $\bm{W}_l \in \mathbb{R}^{d_{l-1} \times d_l}$, and $\bm{b}_l \in \mathbb{R}^{d_l}$ are the output, the weight matrix and bias vector of the $l$-th layer, respectively, and $\sigma$ is the activation function. The input vector $\bm{X} \in \mathbb{R}^s$ of dimension $s$ corresponds to the input $\bm{\phi}_0$, with $d_0 = s$. The number of neurons in each layer is denoted by $d_l$, and $N_l$ is the total number of layers. We adopt the Rectified Linear Unit (ReLU) function \citep{glorot2011deep} $\mathrm{ReLU}(x) = \max(0,x)$ as the activation function $\sigma(x)$. We do not apply an activation function in the final layer of the branch net. In our setup, both the branch net and the trunk net consist of 10 fully-connected layers, each with 128 neurons (i.e., $N_l=10$ and $d_l=128$ for $1 \leq l \leq N_l$).

The outputs of the branch net and the trunk net are vectors, denoted by $\bm{B}_\mathrm{net}$ and $\bm{T}_\mathrm{net}$, respectively. The final output of DeepONet is a scalar $Y$ given by:
\begin{align}
    Y = \bm{B}_\mathrm{net} \cdot \bm{T}_\mathrm{net} + b,
    \label{eq:DONoutput}
\end{align}
where $b \in \mathbb{R}$ is a learnable bias parameter. 

This structure enables the DeepONet architecture to reduce generalization error by distinguishing between physically distinct variables, in contrast to a standard fully-connected NN that treats all inputs as a single, undifferentiated vector \citep{luLearning2021}.

The branch net receives the hydrogen number density $n_\mathrm{H}$, the initial temperature $T_0$, and the initial chemical abundances $\bm{y_0}$ as inputs in the form of $\log(n_\mathrm{H}/\mathrm{cm}^{-3})$, $\log(T_0/\mathrm{K})$, and $\log(\bm{y}_0)$, respectively. The trunk net takes the timestep $\Delta t$ in the form of $\log(1 + \Delta t/\mathrm{sec})$. The output $Y$ is either the temperature ($\log(T/\mathrm{K})$) or the abundance of the $i$-th chemical species ($\log(y(i))$) after evolution over $\Delta t$ from the initial state. We train a separate DeepONet model for each target variable, resulting in a total of seven models for the temperature and the six chemical species.

We define the total loss function as
\begin{align}
    L = \mathrm{MSE} + L1 + L2,
    \label{eq:L_DON}
\end{align}
and train the DeepONet models to minimize this function. Here, the Mean Squared Error (MSE) is defined as
\begin{align}
    \mathrm{MSE} = \frac{1}{N_\mathrm{batch}} \sum_{n=1}^{N_\mathrm{batch}} \left| Y_{n,\mathrm{true}} - Y_{n,\mathrm{pred}} \right|^2,
    \label{eq:MSE_DON}
\end{align}
where $Y_{n,\mathrm{true}}$ and $Y_{n,\mathrm{pred}}$ are the true results obtained from numerical integration and DeepONet predictions, respectively, for the $n$-th training sample, and $N_\mathrm{batch}= 10^5$ denotes the batch size used to divide the training data. To prevent overfitting, we also include both L1 and L2 regularization terms \citep{tibshiraniRegression1996a, hoerl1970ridge1, hoerl1970ridge2}:
\begin{align}
    L1 = \lambda_1 \sum_i |w_i|,\quad L2 = \lambda_2 \sum_i w_i^2, \label{eq:L1L2}
\end{align}
where $w_i$ denotes each element of the weight matrices and bias vectors in Eq.\eqref{eq:one_layer} and the bias parameter in Eq.\eqref{eq:DONoutput}, and the regularization coefficients are set to $\lambda_1 = \lambda_2 = 10^{-8}$.

The DeepONet models are trained using the Adam optimizer \citep{kingmaAdam2014}. During each epoch, which corresponds to training over the entire dataset of $5.12\times10^8$ samples (see Section~\ref{sec:dataset-generation}), the model parameters are updated after every batch of $N_\mathrm{batch}=10^5$ samples, resulting in 5120 iterations per epoch. We start the training with an initial learning rate of $10^{-3}$. The loss function is evaluated every 640 iterations, and the learning rate is halved if no improvement is observed for ten consecutive evaluations (i.e., after 6400 iterations). Note that 640 iterations are much shorter than one epoch. Through multiple trials with different learning-rate scheduling, we empirically found that frequent monitoring is required to achieve stable training.

The range of densities considered is $10^{-3}\,\mathrm{cm^{-3}}\leq n_\mathrm{H}\leq10^{18}\,\mathrm{cm^{-3}}$ and spans more than 20 orders of magnitude. As a single DeepONet model cannot learn the entire range with sufficient accuracy, we partition it into five subranges: $10^{-3}$–$10^4$, $10^4$–$10^8$, $10^8$–$10^{12}$, $10^{12}$–$10^{15}$, and $10^{15}$–$10^{18}\,\mathrm{cm^{-3}}$; hereafter, Regions 0, 1, 2, 3, and 4, respectively. We construct a separate DeepONet model for each of these regions. 

%%%%%%%%%%%%%%%%%%%%%%%%%%%%%%%%%%%%%%%%%%%%%%%%%%%%%%%%%%%%%%%%%%%%%%%%%%
%%%%%%%%%%%%%%%%%%%%%%%%%%%%%%%%%%%%%%%%%%%%%%%%%%%%%%%%%%%%%%%%%%%%%%%%%%

\subsection{Timescale-based Method for Robust Iterative Predictions}
\label{section:timescale-based_method}
\subsubsection{Basic scheme}

Time integration with iterative NN updates can produce unphysical solutions, for instance, through the accumulation of prediction errors \citep{borBridging2025}. To mitigate this, we introduce a simple yet effective method that ensures robust predictions under iterative updates.

Before presenting our method, we briefly explain the mechanism that leads to unphysical solutions. Consider the evolution of a variable under repeated NN predictions. If the timestep is very short, the physical change over that step can be smaller than the prediction error. In such a case, the time variation is dominated by the error rather than by the physical evolution, and iterating such updates drives the solution away from the physical one.

To address this issue, we propose a modified update method based on a characteristic timescale for significant variation. For a physical quantity $x$, we choose a dimensionless parameter $\epsilon$ significantly smaller than the typical relative error of DeepONet predictions. Then, we define the characteristic timescale $\tau_\epsilon$ as the interval over which the physical evolution changes $x$ by a fraction $\epsilon$, i.e., $|\Delta x(\tau_\epsilon)|=\epsilon\,|x|$.

Finally, we define the change in a quantity $x$ over a timestep $\Delta t$ as
\begin{align}
\Delta x(\Delta t)=
\begin{cases}
\dfrac{\Delta t}{\tau_\epsilon}\,\Delta x_{\mathrm{pred}}(\tau_\epsilon), & \Delta t<\tau_\epsilon,\\[6pt]
\Delta x_{\mathrm{pred}}(\Delta t), & \Delta t\ge\tau_\epsilon,
\end{cases}
\label{eq:timescale-based}
\end{align}
where $\Delta x_{\mathrm{pred}}(\Delta t')$ is the change predicted by DeepONet over an interval $\Delta t'$. Hereafter, we call this method the \textit{timescale-based method}.

When $\Delta t < \tau_\epsilon$, this method replaces the short-timestep prediction $\Delta x_{\mathrm{pred}}(\Delta t)$, dominated by the prediction error, with a scaled version of a longer-timestep prediction $(\Delta t/\tau_\epsilon)\,\Delta x_{\mathrm{pred}}(\tau_\epsilon)$, dominated by physical evolution. On the other hand, when $\Delta t \ge \tau_\epsilon$, the method simply returns the DeepONet prediction $\Delta x_{\mathrm{pred}}(\Delta t)$ because the prediction error is subdominant. As a result, the prediction error is always smaller than the physical evolution at each update, enabling robust predictions even under short-timestep iterations.

In Eq.~\eqref{eq:timescale-based}, we cap the value of $\tau_\epsilon$ at the maximum timestep learned by DeepONet, $\Delta t_\mathrm{max}$, since NNs generally make unreliable predictions outside the learned region. In addition, we enforce the hydrogen nucleus conservation and charge neutrality conditions after each update to improve the stability of evolution.

In general, the characteristic timescale $\tau_\epsilon$ differs for each target variable. Since our DeepONet framework uses a separate model for each variable, each variable is updated with its own DeepONet model and  $\tau_\epsilon$ in Eq.~\eqref{eq:timescale-based}.

%%%%%%%%%%%%%%%%%%%%%%%%%%%%%%%%%%%%%%%%%%%%%%%%%%%%%%%%%%%%%%%%%%%%%%%%%%

\subsubsection{Neural Network for Estimating Characteristic Timescales: Deep Galerkin Method}
\label{sec:deep-galerkin-method}

Naively, one may think $\tau_\epsilon$ can be obtained by using the instantaneous time derivative $\dot{x}$ as $\tau_\epsilon = \epsilon x / \dot{x}$. However, due to the stiffness of the system, $\tau_\epsilon$ obtained in this way can be orders-of-magnitude smaller than the actual time required for the fractional change of $x$ to reach $\epsilon$. Alternatively, $\tau_\epsilon$ can be obtained by numerical integration of the ODEs until the change in $x$ reaches $\pm\epsilon x$. However, such a procedure is computationally expensive. 

To reduce the computational cost of estimating $\tau_\epsilon$, we introduce new NNs in addition to the DeepONet models described in Section~\ref{sec:deeponet}. Here, we construct Deep Galerkin Method (DGM)-based NNs \citep{sirignanoDGM2018} to estimate $\tau_\epsilon$, using PyTorch \citep{paszkePyTorch2019}. The DGM architecture, which has been applied to modeling the ISM \citep{brancaNeural2023}, is suited to problems where the output changes sharply in response to the input, such as estimating $\tau_\epsilon$ for a given chemical and thermal state.

Following \citet{brancaNeural2023}, the architecture of the DGM network is defined as \citep[see also][]{sirignanoDGM2018}:
\begin{align}
    \bm{\phi}_0 &= \sigma\left(\bm{X}\bm{U}_0 + \bm{b}_0\right), \label{eq:dgm-0}\\
    \bm{Z}_l &= \sigma\left(\bm{X}\bm{U}_l^{(z)} + \bm{\phi}_{l-1}\bm{W}_l^{(z)} + \bm{b}_l^{(z)}\right), \label{eq:dgm-1}\\
    \bm{G}_l &= \sigma\left(\bm{X}\bm{U}_l^{(g)} + \bm{\phi}_{l-1}\bm{W}_l^{(g)} + \bm{b}_l^{(g)}\right), \label{eq:dgm-2}\\
    \bm{R}_l &= \sigma\left(\bm{X}\bm{U}_l^{(r)} + \bm{\phi}_{l-1}\bm{W}_l^{(r)} + \bm{b}_l^{(r)}\right), \label{eq:dgm-3}\\
    \bm{H}_l &= \sigma\left(\bm{X}\bm{U}_l^{(h)} + (\bm{\phi}_{l-1} \odot \bm{R}_l)\bm{W}_l^{(h)} + \bm{b}_l^{(h)}\right), \label{eq:dgm-4}\\
    \bm{\phi}_l &= (1 - \bm{G}_l) \odot \bm{H}_l + \bm{Z}_l \odot \bm{\phi}_{l-1}, \label{eq:dgm-5}\\
    \bm{Y} &= \bm{\phi}_{N_l} \bm{W} + \bm{b}, \label{eq:dgm-6}
\end{align}
where the intermediate DGM layers ($l=1,2,\dots,N_l$ with $N_l$ the number of DGM layers) are sandwiched by the fully-connected first and last layers. Each DGM layer receives both the input vector $\bm{X} \in \mathbb{R}^s$ and the output of the previous layer $\bm{\phi}_{l-1}$. At each DGM layer, intermediate states $\bm{Z}_l$, $\bm{G}_l$, $\bm{R}_l$, and $\bm{H}_l\in\mathbb{R}^{d_l}$ are computed and then combined to yield the layer output $\bm{\phi}_{l}\in\mathbb{R}^{d_l}$ (here, $\odot$ denotes the Hadamard product). For each of ${(*)}\in \{z,g,r,h\}$,  $\bm{U}_l^{(*)}\in\mathbb{R}^{s \times d_l}$, $\bm{W}_l^{(*)} \in\mathbb{R}^{d_{l-1}\times d_l}$, $\bm{b}_l^{(*)}\in\mathbb{R}^{d_l}$ are the weights for $\bm{X}$, the weights for $\bm{\phi}_{l-1}$, and the biases, respectively. Finally, the last layer yields the output vector $\bm{Y}$, without applying any activation function. Unlike conventional fully-connected layers, each DGM layer consists of repeated element-wise multiplications involving nonlinear transformations of the input, making DGM networks suited to emulating complex and rapidly varying functions \citep{sirignanoDGM2018}.

The input is the same as that of the branch net of DeepONet (Section~\ref{sec:deeponet}), i.e, a vector of the density and the initial temperature and abundances in the form of $\log(n_\mathrm{H}/\mathrm{cm}^{-3})$, $\log(T_0/\mathrm{K})$, and $\log(\bm{y}_0)$. The output is a vector of the timescales $\tau_\epsilon$ for the temperature and six chemical abundances. Unlike DeepONet, a single DGM model outputs all target variables simultaneously. We choose the sigmoid function $\mathrm{sigmoid}(x) = (1 + e^{-x})^{-1}$ as the activation function $\sigma(x)$. In our setup, the DGM network consists of 10 DGM layers, each with 128 neurons.

To facilitate learning, instead of estimating $\tau_\epsilon$ directly, we introduce the auxiliary variable
\begin{align}
Y_\tau =
\begin{cases}
\log\!\left(1 + \epsilon\,\dfrac{t_\mathrm{ff}}{\tau_\epsilon}\right), & \Delta x > 0,\\[6pt]
-\log\!\left(1 + \epsilon\,\dfrac{t_\mathrm{ff}}{\tau_\epsilon}\right), & \Delta x < 0.
\end{cases}
\label{eq:Y_tau}
\end{align}
For $\epsilon\, t_\mathrm{ff}/\tau_\epsilon\ll1$ we have $Y_\tau\simeq\pm\epsilon\, t_\mathrm{ff}/\tau_\epsilon$, corresponding to the normalized time derivative $\dot{x}/x$. When $\epsilon\, t_\mathrm{ff}/\tau_\epsilon\gg1$, the logarithmic form in Eq.~\eqref{eq:Y_tau} bounds $Y_\tau$ and avoids numerical difficulties caused by very large $1/\tau_\epsilon$.

For the DGM network, we extend the loss function of DeepONet (Eq.~\eqref{eq:L_DON}) to handle multiple outputs. Specifically, we modify the MSE of DeepONet (Eq.~\eqref{eq:MSE_DON}) by summing over all output components:
\begin{align}
    \mathrm{MSE}_\tau = \sum_{i=0}^6 \left( \frac{1}{N_\mathrm{batch}} \sum_{n=1}^{N_\mathrm{batch}} \left| Y_{\tau,i,n,\mathrm{true}} - Y_{\tau,i,n,\mathrm{pred}} \right|^2 \right), \label{eq:MSE_tau}
\end{align}
where $i$ corresponds to the seven components (temperature and six chemical species), and $Y_{\tau,i,n,\mathrm{true}}$ and $Y_{\tau,i,n,\mathrm{pred}}$ denote the true and estimated values of $Y_\tau$ for the $i$-th component for the $n$-th training sample, respectively. In DGM network training, we also take $N_\mathrm{batch}=10^5$. We add L1 and L2 regularization (Eq.~\eqref{eq:L1L2}) on all weights and biases, with $\lambda_1 = 10^{-6}$ and $\lambda_2 = 10^{-7}$. The total loss is then:
\begin{align}
    L_\tau = \mathrm{MSE}_\tau + L1 + L2.
\end{align}
Finally, we split the density range into the same five regions used for DeepONet (Section~\ref{sec:deeponet}) and train a separate DGM model in each region.

%%%%%%%%%%%%%%%%%%%%%%%%%%%%%%%%%%%%%%%%%%%%%%%%%%%%%%%%%%%%%%%%%%%%%%%%%%
%%%%%%%%%%%%%%%%%%%%%%%%%%%%%%%%%%%%%%%%%%%%%%%%%%%%%%%%%%%%%%%%%%%%%%%%%%

\subsection{Dataset Generation}
\label{sec:dataset-generation}

Here, we explain how we generate training datasets for the two types of NNs, the DeepONet (Section~\ref{sec:deeponet}) and DGM (Section~\ref{sec:deep-galerkin-method}) models. The ranges of temperature and chemical abundances in the datasets are designed to sufficiently cover the one-zone thermochemical evolution paths for Pop III star formation (Section~\ref{sec:basic-equations}).

We generate datasets by numerically integrating the thermochemical evolution equations (Eqs.~\eqref{ydot} and \eqref{Tdot}) from a given initial state. To ensure numerical stability and accuracy, the equations are integrated using a fully-implicit scheme with adaptive substepping. Although the fully-implicit scheme is numerically stable, the errors can become non-negligible when the variations in $T$ or $y(i)$ are large. Therefore, the timestep is subdivided so that the relative changes in $T$ and $y(i)$ remain within 10\,\% per substep. We have confirmed that further reduction of the substep size (below 10\,\%) does not significantly affect the results. Hence, we adopt the 10\% substepping criterion throughout this work.

For both the DeepONet and DGM models, the ranges of the input parameters for each density region (see Section~\ref{sec:deeponet}) are summarized in Table~\ref{table:RegionIC}. To generate initial states, we first sample the temperature and chemical abundances uniformly within the ranges shown in Table~\ref{table:RegionIC}. We then adjust the abundances to enforce hydrogen-nucleus conservation and overall charge neutrality. Because this procedure can shift the abundances outside the original parameter space, primarily toward higher values, the table should be understood as defining the original sampling range, while the actual parameter space may be somewhat broader.

For the DeepONet models, we also randomly sample the timestep parameter $\log(1+\Delta t/\mathrm{sec})$ between $0$ and $\log\!\bigl(1+\Delta t_{\max}/\mathrm{sec}\bigr)$, with $\Delta t_{\max}=10^{0.5}\,t_\mathrm{ff}$, and evolve each initial state for $\Delta t$. Each training dataset contains $5.12 \times 10^8$ samples. For validation, an additional set of 5,120 samples is created. Generating the dataset required roughly $200$\,CPUhrs per region.

For the DGM models, given that the prediction errors are typically less than 5\% (or mostly less than 10\%; see Table~\ref{table:error} in Section~\ref{section:error_main} below), we compute $\tau_\epsilon$ for $\epsilon = 0.1$. We have confirmed that the exact value of $\epsilon$, and hence $\tau_\epsilon$, does not affect the results significantly by performing test computations with doubled $\tau_\epsilon$. In obtaining $\tau_\epsilon$, we evolve each initial state while recording the time when the relative change of each variable reaches $10\,\%$. The size of the training dataset and an additional validation dataset for each region is again $5.12 \times 10^8$ and 5,120 samples, respectively. The dataset generation time for each region was approximately 170 CPUhrs. 

Training was conducted on NVIDIA A100 GPUs (40 GB). The DeepONet models were trained for 50 epochs, requiring about 1.67 GPUhrs per variable (temperature or abundance), leading to a total of 11.7 GPUhrs per region. The DGM models were trained for 100 epochs and required approximately 15.2 GPUhrs per region.

\begin{deluxetable*}{lccccccccccccccc}
\tablewidth{0pt}
\tablecaption{Parameter ranges for training dataset in each density region \label{table:RegionIC}}
\tablehead{
    && \multicolumn{2}{c}{Region 0} && \multicolumn{2}{c}{Region 1} && \multicolumn{2}{c}{Region 2} && \multicolumn{2}{c}{Region 3} && \multicolumn{2}{c}{Region 4} \\
    \tableline
    && min & max && min & max && min & max && min & max && min & max 
}
\startdata
    $\log(n_{\mathrm{H}}/\mathrm{cm}^{-3})$ && $-3$  & $4$     && $4$   & $8$     && $8$   & $12$    && $12$  & $15$    && $15$  & $18$   \\
    $\log(T/\mathrm{K})$                    && $1.3$ & $4.5$   && $1.3$ & $4.5$   && $1.3$ & $4.0$   && $1.3$ & $4.0$   && $1.3$ & $4.0$  \\
    $\log(y(\mathrm{H}))$                   && $-6$  & $0$     && $-6$  & $0$     && $-6$  & $0$     && $-6$  & $0$     && $-6$  & $0$    \\
    $\log(y(\mathrm{H}_2))$                 && $-8$  & $-0.3$  && $-6$  & $-0.3$  && $-6$  & $-0.3$  && $-6$  & $-0.3$  && $-6$  & $-0.3$ \\
    $\log(y(\mathrm{e^-}))$                 && $-10$ & $-4$    && $-10$ & $-4$    && $-12$ & $-6$    && $-14$ & $-8$    && $-15$ & $-9$   \\
    $\log(y(\mathrm{H}^+))$                 && $-10$ & $-4$    && $-10$ & $-4$    && $-12$ & $-6$    && $-14$ & $-8$    && $-15$ & $-9$   \\
    $\log(y(\mathrm{H}^-))$                 && $-19$ & $-13$   && $-20$ & $-14$   && $-21$ & $-15$   && $-21$ & $-15$   && $-22$ & $-16$  \\
    $\log(y(\mathrm{H}_2^+))$               && $-19$ & $-13$   && $-20$ & $-14$   && $-21$ & $-15$   && $-21$ & $-15$   && $-22$ & $-16$  \\
\enddata
\tablecomments{
The value of $1.3$ for the lower bound of temperature corresponds to $\log(20)$, i.e., a temperature of $T = 20$ K. The value of $-0.3$ for the upper bound of H$_2$ abundance corresponds to $\log(0.5)$, which reflects the constraint that $y(\mathrm{H}_2)$ cannot exceed 0.5.
}
\end{deluxetable*}

\begin{figure}[htb!]
    \includegraphics[width=9cm]{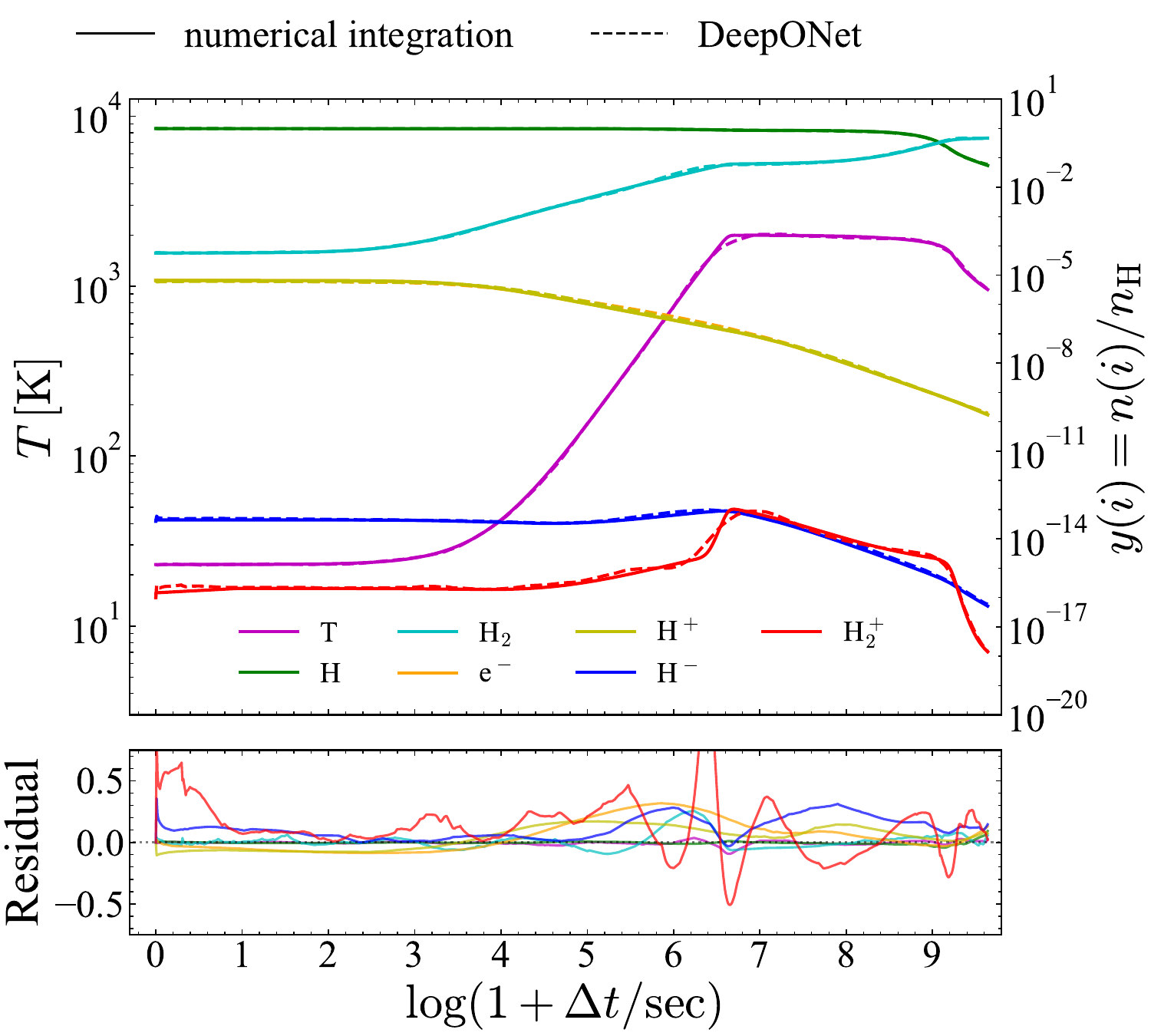}
\caption{
Single-step DeepONet predictions of the thermal and chemical evolution for a fixed hydrogen number density of $n_\mathrm{H} = 10^{12}\,\mathrm{cm}^{-3}$. The top panel compares the DeepONet predictions (dashed) with the ground-truth results from numerical integrations (solid). The bottom panel plots the corresponding residual errors normalized by the ground-truth values, i.e., $\left(T_\mathrm{pred}-T_\mathrm{true}\right)/T_\mathrm{true}$ or $\left(y_\mathrm{pred}(i)-y_\mathrm{true}(i)\right)/y_\mathrm{true}(i)$, where the subscripts ``pred'' and ``true'' denote the DeepONet and ground-truth values, respectively. The horizontal axis represents the transformed timestep, $\log(1 + \Delta t/\mathrm{sec})$. The DeepONet predictions almost perfectly reproduce the ground-truth results. The variables are color-coded as follows: $T$ (magenta); $y(\mathrm{H})$ (green); $y(\mathrm{H}_2)$ (cyan); $y(\mathrm{e}^-)$ (orange); $y(\mathrm{H}^+)$ (yellow); $y(\mathrm{H}^-)$ (blue); and $y(\mathrm{H}_2^+)$ (red). Note that the curves for $y(\mathrm{e}^-)$ are nearly invisible because they almost completely coincide with those for $y(\mathrm{H}^+)$.
}
    \label{fig:single_cell_nocomp}
\end{figure}

\begin{figure*}[htb!]
    \includegraphics[width=17cm]{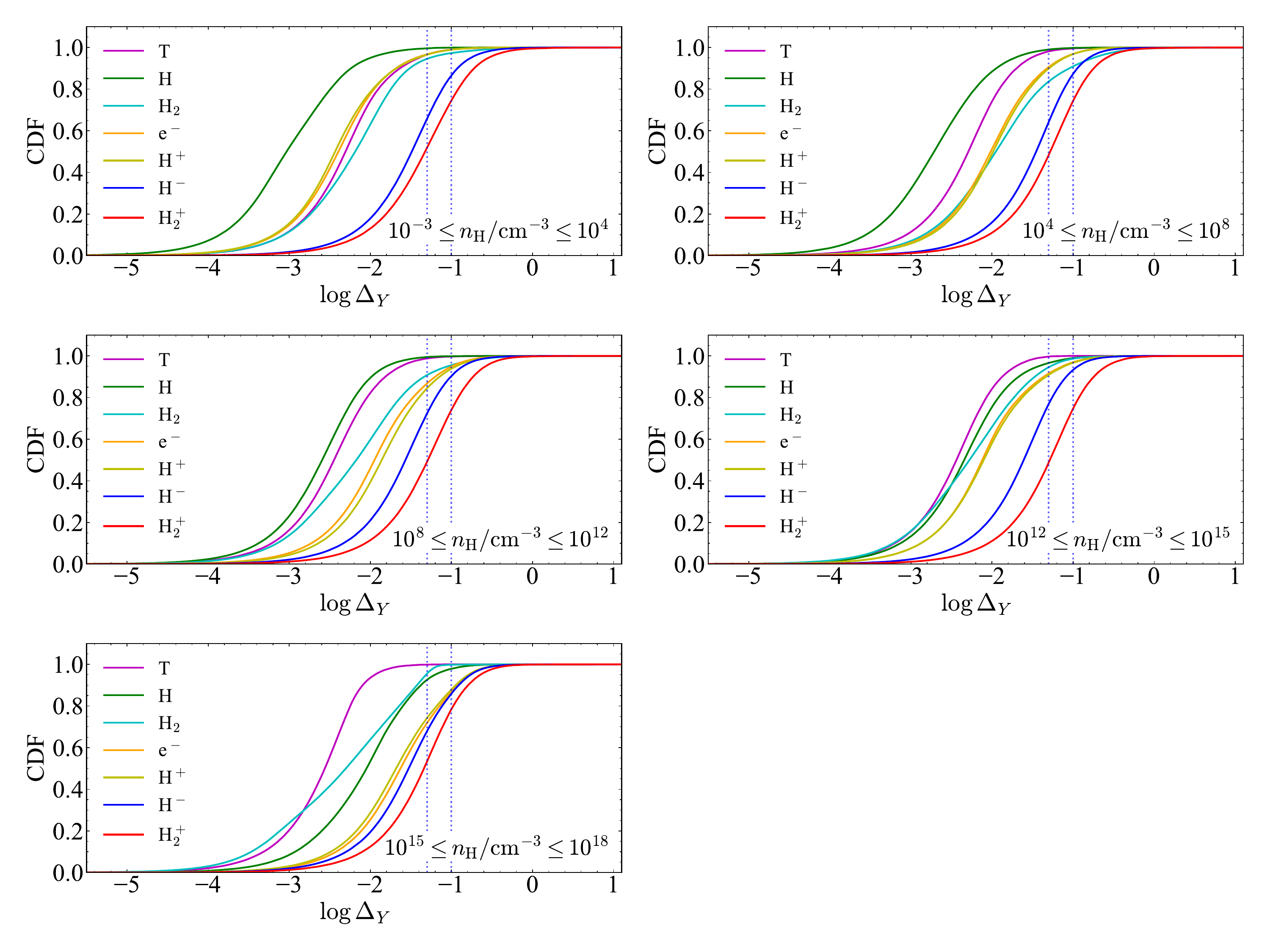}
\caption{
Cumulative distribution functions of the relative prediction errors of the DeepONet emulator for each density region. The horizontal axis shows the logarithm of relative error $\Delta_Y$, defined in Eq.~(\ref{eq:delta_Y}), and the vertical axis indicates the cumulative relative frequency. The two blue dotted vertical lines show $\Delta_Y=0.05$ (left) and $0.1$ (right). Each panel corresponds to a different density range: Region 0 ($10^{-3}\leq n_\mathrm{H}/\mathrm{cm^{-3}}\leq10^4$, top left); Region 1 ($10^4\leq n_\mathrm{H}/\mathrm{cm^{-3}}\leq10^8$, top right); Region 2 ($10^8\leq n_\mathrm{H}/\mathrm{cm^{-3}}\leq10^{12}$, middle left); Region 3 ($10^{12}\leq n_\mathrm{H}/\mathrm{cm^{-3}}\leq10^{15}$, middle right); and Region 4 ($10^{15}\leq n_\mathrm{H}/\mathrm{cm^{-3}}\leq10^{18}$, bottom left). Colors are the same as in Figure~\ref{fig:single_cell_nocomp}.
}   \label{fig:single_cell_CDF}
\end{figure*}

\begin{deluxetable*}{lrrrrrrrrrrrrrrr}
\tablewidth{0pt}
\tablecaption{
Fraction of predictions with relative errors $\Delta_Y$ below 5\% and 10\% for each variable in each density region.
\label{table:error}
}
\tablehead{&& \multicolumn{2}{c}{Region 0} && \multicolumn{2}{c}{Region 1} && \multicolumn{2}{c}{Region 2} && \multicolumn{2}{c}{Region 3} && \multicolumn{2}{c}{Region 4}\\
\tableline
$\Delta_Y$ && $<5\%$ & $<10\%$ && $<5\%$ & $<10\%$ && $<5\%$ & $<10\%$ && $<5\%$ & $<10\%$ && $<5\%$ & $<10\%$
}
\startdata
    $T$              && $97\%$  & $99\%$  && $98\%$ & $100\%$ && $99\%$ & $100\%$ && $100\%$ & $100\%$ && $100\%$ & $100\%$  \\
    $\mathrm{H}$     && $100\%$ & $100\%$ && $99\%$ & $100\%$ && $99\%$ & $100\%$ && $97\%$  & $99\%$  && $92\%$  & $98\%$  \\
    $\mathrm{H}_2$   && $94\%$  & $97\%$  && $84\%$ & $91\%$  && $91\%$ & $96\%$  && $94\%$  & $99\%$  && $95\%$  & $100\%$  \\
    $\mathrm{e}^-$   && $97\%$  & $99\%$  && $91\%$ & $97\%$  && $87\%$ & $95\%$  && $91\%$  & $97\%$  && $72\%$  & $87\%$  \\
    $\mathrm{H}^+$   && $96\%$  & $99\%$  && $90\%$ & $97\%$  && $84\%$ & $94\%$  && $91\%$  & $97\%$  && $74\%$  & $88\%$  \\
    $\mathrm{H}^-$   && $66\%$  & $86\%$  && $64\%$ & $87\%$  && $72\%$ & $90\%$  && $77\%$  & $93\%$  && $68\%$  & $86\%$  \\
    $\mathrm{H}_2^+$ && $52\%$  & $74\%$  && $48\%$ & $74\%$  && $49\%$ & $74\%$  && $49\%$  & $75\%$  && $53\%$  & $78\%$  \\
\enddata
\tablecomments{
Each density region is defined as follows: Region 0 ($10^{-3}\leq n_\mathrm{H}/\mathrm{cm^{-3}}\leq10^4$); Region 1 ($10^4\leq n_\mathrm{H}/\mathrm{cm^{-3}}\leq10^8$); Region 2 ($10^8\leq n_\mathrm{H}/\mathrm{cm^{-3}}\leq10^{12}$); Region 3 ($10^{12}\leq n_\mathrm{H}/\mathrm{cm^{-3}}\leq10^{15}$); and Region 4 ($10^{15}\leq n_\mathrm{H}/\mathrm{cm^{-3}}\leq10^{18}$).
}
\end{deluxetable*}

\begin{figure}[htb!]
 \begin{center}
   \includegraphics[width=9cm]{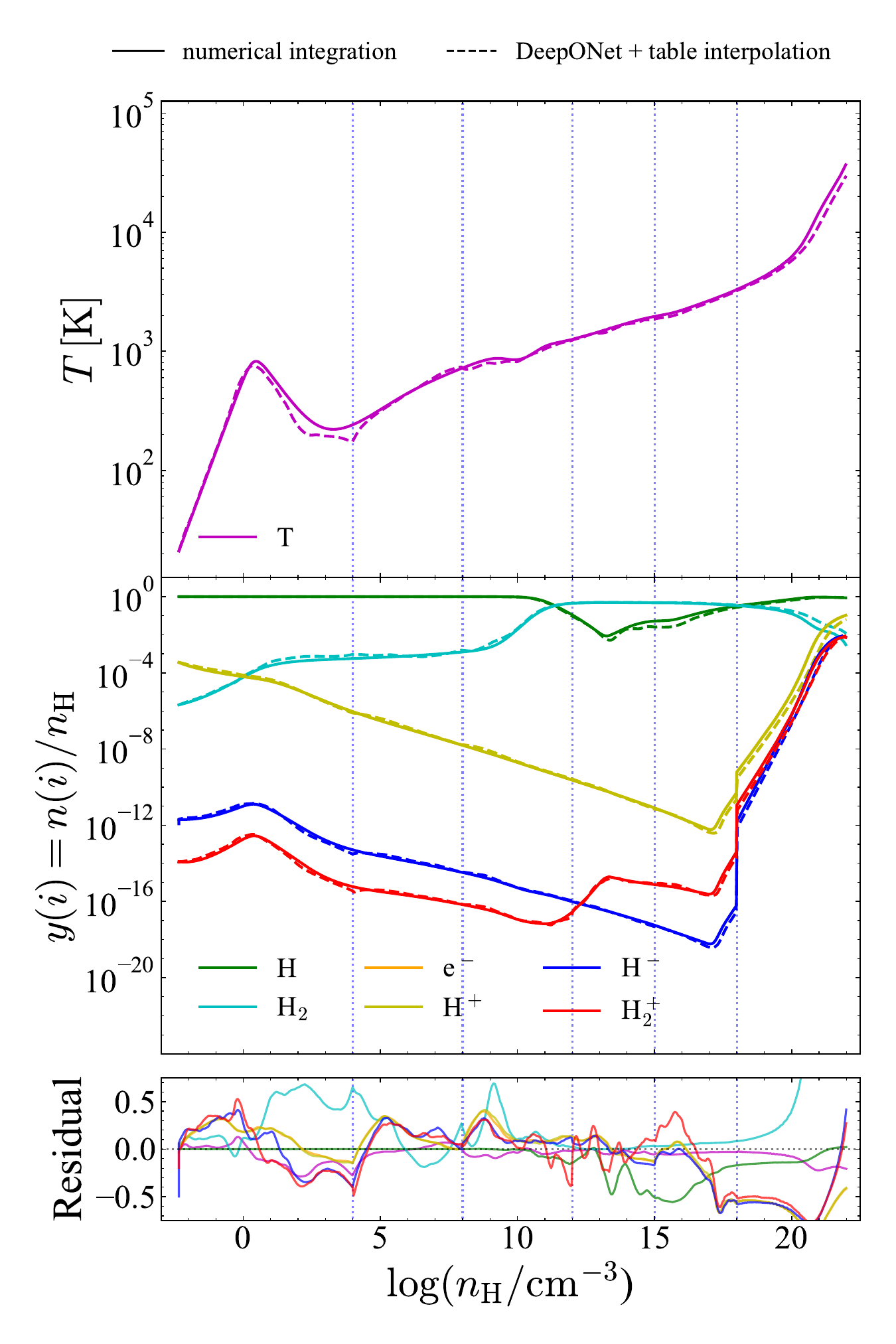}
 \end{center}
\caption{
One-zone evolution with timestep $\Delta t = 10^{-1}\,t_\mathrm{ff}$. The horizontal axis is the hydrogen number density $n_\mathrm{H}$, corresponding to the time evolution during gravitational collapse. The top panel shows the temperature evolution, the middle panel shows the chemical evolution, and the bottom panel shows the residual errors. Blue dotted vertical lines indicate the boundaries between different density regions. Colors are the same as in Figure~\ref{fig:single_cell_nocomp}. The solid lines show the ground-truth numerical integration results, while the dashed lines show the DeepONet predictions for $n_{\mathrm{H}} \leq 10^{18}\,\mathrm{cm^{-3}}$ and the results obtained by interpolating a precalculated table for $n_{\mathrm{H}} \ge 10^{18}\,\mathrm{cm^{-3}}$ (see Appendix~\ref{sec:app-table-equilibrium}). Note that the apparent jump in the abundances of e$^-$, H$^+$, H$^-$, and H$_2^+$ at $n_{\mathrm{H}} = 10^{18}\,\mathrm{cm^{-3}}$, arising due to the absence of inverse reactions in our chemical network, does not affect the thermal evolution \citep{Nakauchi:2019}. The abundances of $\mathrm{e}^-$ and $\mathrm{H}^+$ overlap almost completely with each other.
}\label{fig:onezone_nocomp}
\end{figure}

\begin{figure*}[htb!]
    \includegraphics[width=17cm]{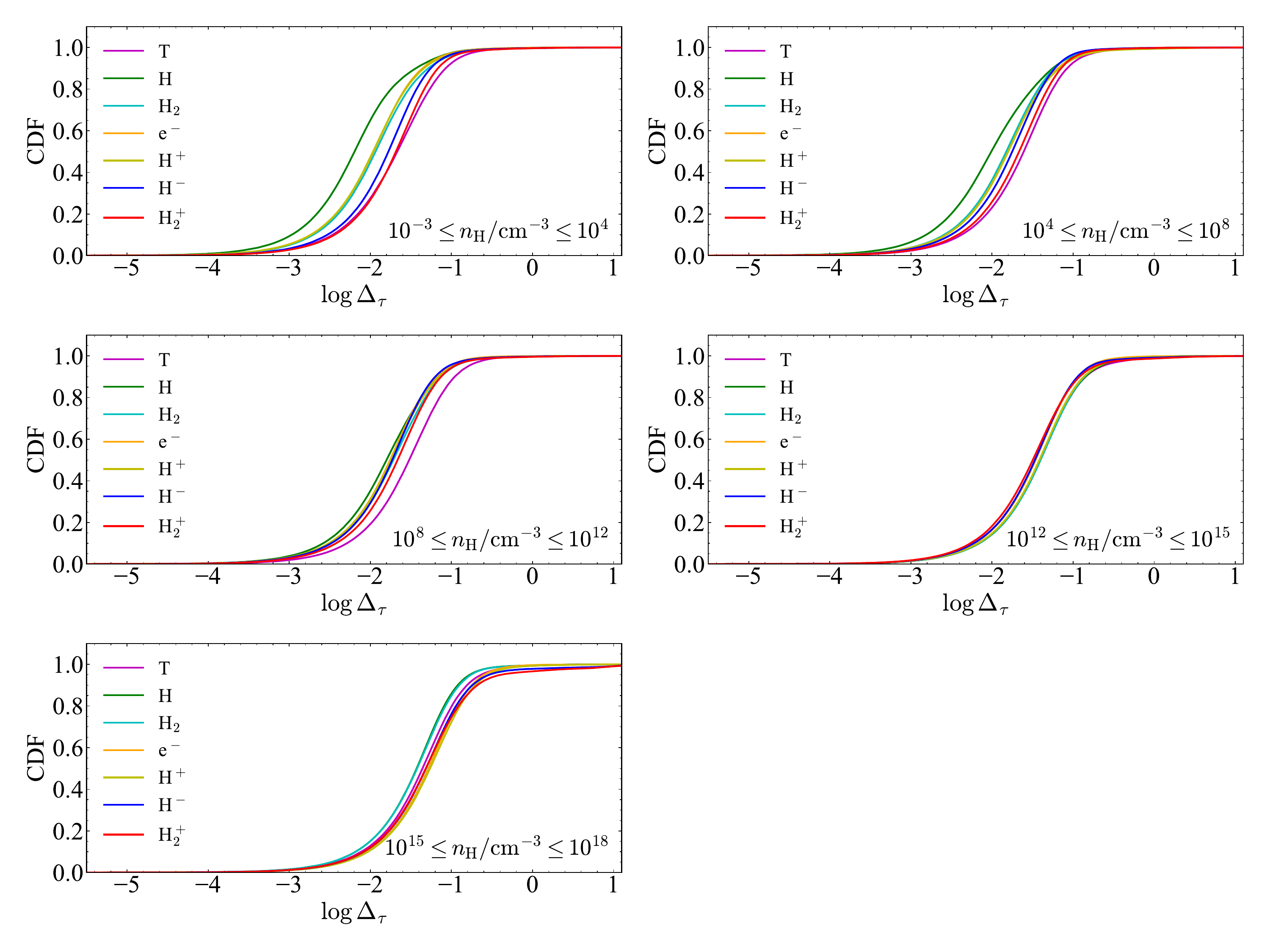}
\caption{
Same as Figure~\ref{fig:single_cell_CDF}, but for the characteristic timescales $\tau_\epsilon$ with $\epsilon=0.1$. The horizontal axis shows the logarithm of $\Delta_\tau$ as defined in Eq.~\eqref{eq:delta_chemcool}.
}   \label{fig:t_chemcool_CDF}
\end{figure*}

\begin{figure*}[htb!]
 \begin{center}
   \includegraphics[width=18cm]{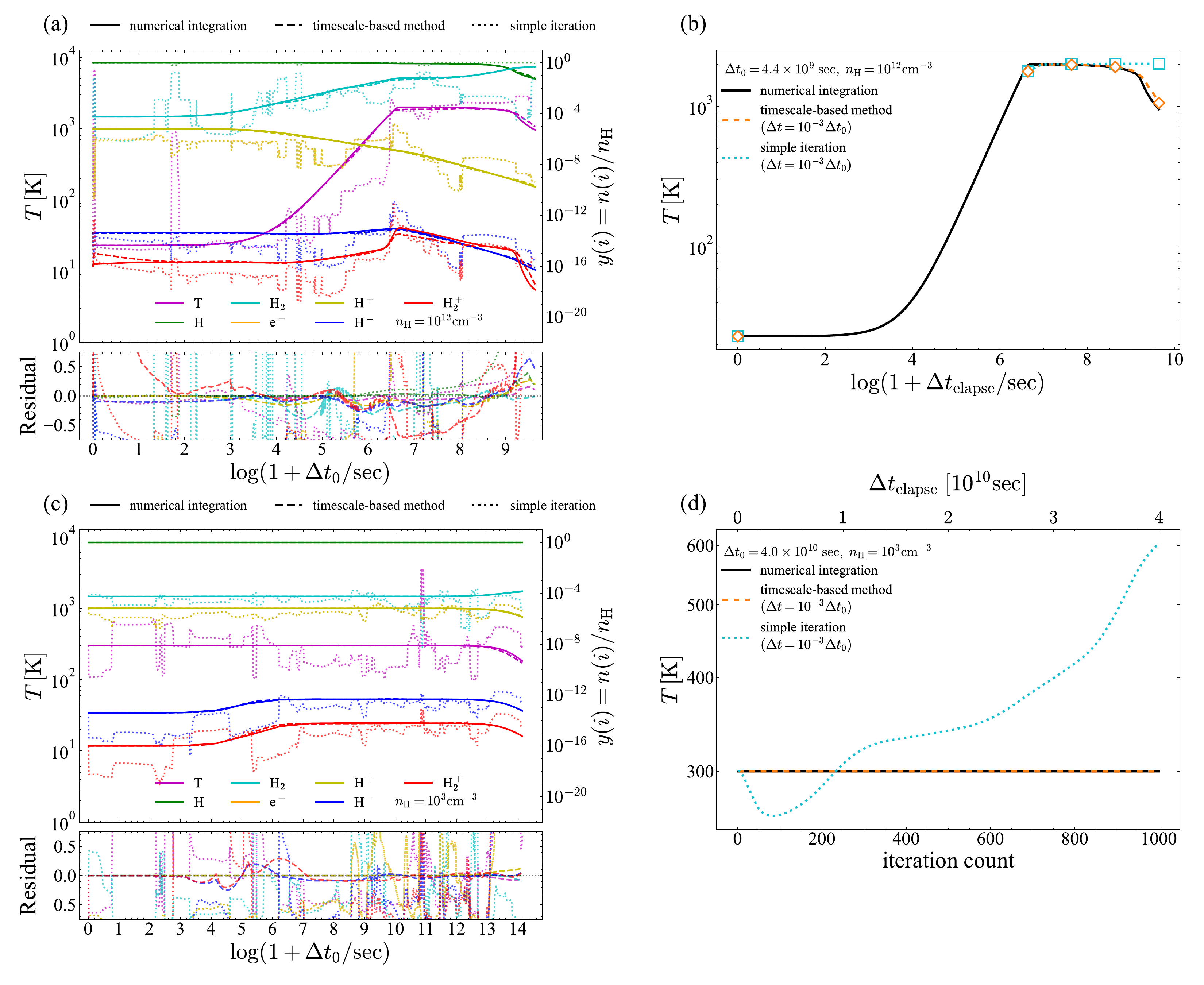}
 \end{center}
\caption{
Comparison of fixed-density evolution using simple iteration of DeepONet predictions (dotted), the timescale-based method (dashed), and numerical integration (solid). (a) Same as Figure~\ref{fig:single_cell_nocomp} but for iterative updates. The horizontal axis is the overall timestep $\Delta t_{0}$, which is divided into 1000 equal substeps of $\Delta t=\Delta t_{0}/1000$ for iterations. (b) Step-by-step temperature evolution for the case of $\Delta t_{0}=\Delta t_{\max}=4.4\times10^{9}\,\mathrm{sec}$ in panel (a). (c) Same as panel (a) but for the case with the initial temperature $T=300\,\mathrm{K}$ and fixed density $n_{\mathrm{H}}=10^{3}\,\mathrm{cm^{-3}}$. (d) Step-by-step temperature evolution for the case of $\Delta t_{0}=4.0\times10^{10}\,\mathrm{sec}$ in panel (c). In panels (a) and (c), colors are the same as in Figure~\ref{fig:single_cell_nocomp}, and $y(\mathrm{e}^-)$ and $y(\mathrm{H}^+)$ nearly coincide. The bottom panels of (a) and (c) show the residual errors, with line styles corresponding to those in the panels above. In panel (b), the horizontal axis shows the elapsed time $\Delta t_{\mathrm{elapse}}=i\,\Delta t$ at the iteration count $i$, with markers indicating the initial state and the 1st, 10th, 100th, and 1000th iterations. In panel (d), the bottom horizontal axis shows the iteration count, while the top axis indicates $\Delta t_\mathrm{elapse}$.
}\label{fig:fixed_density_merge}
\end{figure*}

\begin{figure}[htb!]
    \includegraphics[width=9cm]{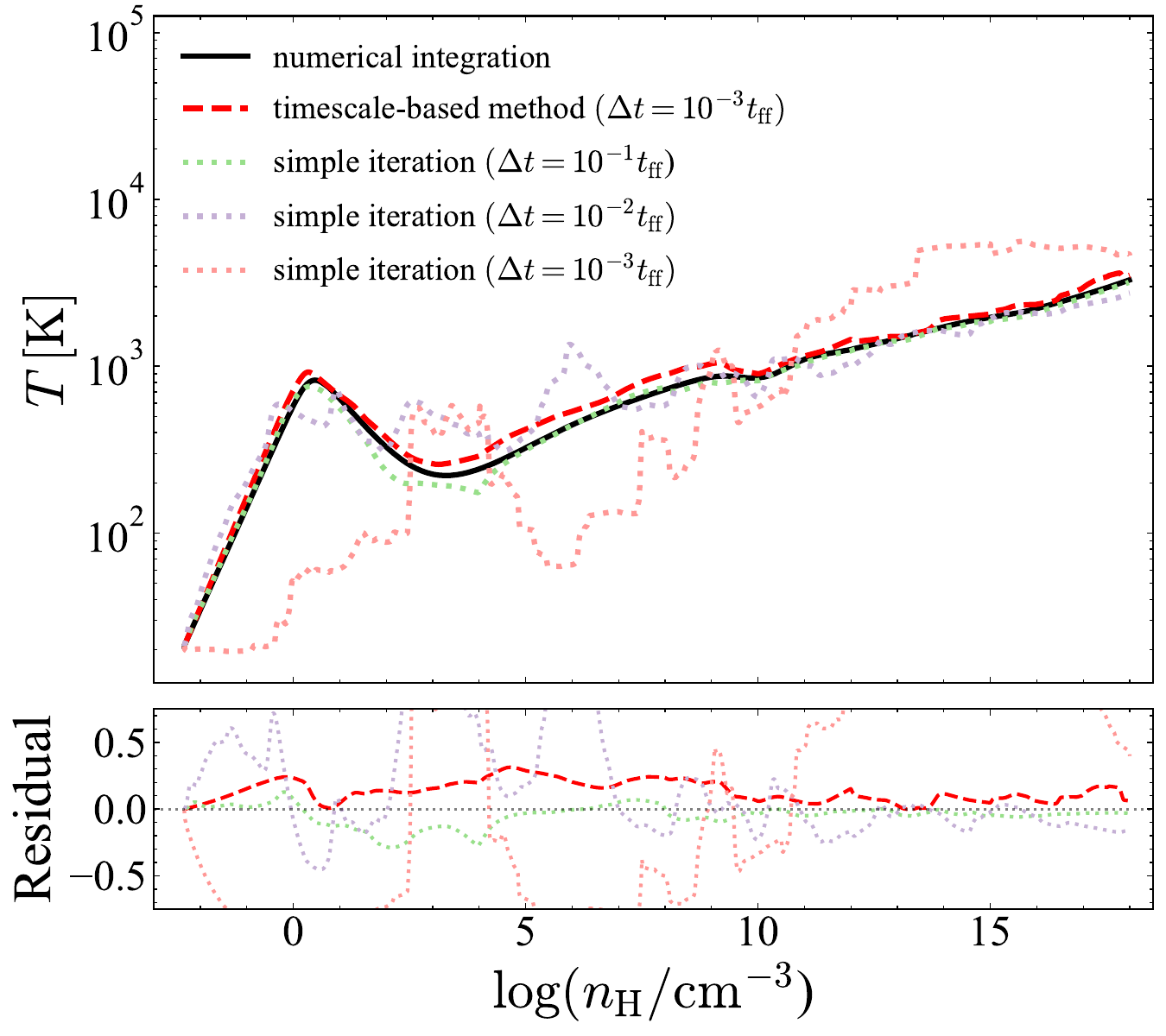}
 \caption{
Same as the top and bottom panels of Figure~\ref{fig:onezone_nocomp}, but with different time steps: $\Delta t = 10^{-1}$ (green), $10^{-2}$ (purple), and $10^{-3} \,t_\mathrm{ff}$ (red). We show the evolution using simple iteration of DeepONet predictions (dotted), the timescale-based method (dashed), and numerical integration (solid). Although only the case with $\Delta t = 10^{-3}\, t_\mathrm{ff}$ is shown for the timescale-based method, the results for different timesteps, such as $10^{-1}$, $10^{-2}$, and $10^{-4}\,t_\mathrm{ff}$, are approximately the same.}
    \label{fig:onezone_complement}
\end{figure}

%%%%%%%%%%%%%%%%%%%%%%%%%%%%%%%%%%%%%%%%%%%%%%%%%%%%%%%%%%%%%%%%%%%%%%%%%%
%%%%%%%%%%%%%%%%%%%%%%%%%%%%%%%%%%%%%%%%%%%%%%%%%%%%%%%%%%%%%%%%%%%%%%%%%%

\section{Results}\label{results}

%%%%%%%%%%%%%%%%%%%%%%%%%%%%%%%%%%%%%%%%%%%%%%%%%%%%%%%%%%%%%%%%%%%%%%%%%%
%%%%%%%%%%%%%%%%%%%%%%%%%%%%%%%%%%%%%%%%%%%%%%%%%%%%%%%%%%%%%%%%%%%%%%%%%%

\subsection{Performance of DeepONet-based Emulator}
\label{sec:neur-netw-pred}
Here, we evaluate the performance of the DeepONet-based emulator (Section~\ref{sec:deeponet}) in two cases: single-step evolution from an initial state (Section~\ref{section:error_main}) and a one-zone collapse calculation (Section~\ref{sec:euml-therm-chem}).  These tests demonstrate that the emulator successfully learns the thermochemical evolution over the wide range of densities relevant to Pop III star formation. 

\subsubsection{Single-step predictions of thermal and chemical evolution at a fixed density}
\label{section:error_main}

Here, we demonstrate that our DeepONet models can accurately predict the thermochemical evolution of primordial gas. We begin with an example of time evolution from a specific initial condition and then assess the distribution of prediction errors for a large sample of initial conditions.

Let us examine the time evolution at a fixed density of $n_{\mathrm{H}} = 10^{12}\,\mathrm{cm}^{-3}$ starting from the following initial condition:  $T = 23$\,K, $y(\mathrm{H}_2) = 5.7 \times 10^{-5}$, $y(\mathrm{H}^+) = 6.6 \times 10^{-6}$, $y(\mathrm{H}^-) = 4.0 \times 10^{-14}$, and $y(\mathrm{H}_2^+) = 1.0 \times 10^{-16}$, with $y(\mathrm{H})$ and $y(\mathrm{e}^-)$ determined to satisfy the hydrogen nucleus conservation and charge neutrality conditions. We examine single-step predictions over the timestep $\Delta t$, sampled logarithmically as $\log(1+\Delta t/\mathrm{sec})=(i/1000)\log(1+\Delta t_{\mathrm{max}}/\mathrm{sec}) $ for $i = 1, 2, \dots, 1000$. Here, we employ the model for Region 3, while the density $n_{\mathrm{H}} = 10^{12}\,\mathrm{cm}^{-3}$ is located at the boundary of Regions 2 and 3 (see Table~\ref{table:RegionIC}).

Note that the initial chemical abundances fall slightly outside the range of Region 3 in Table~\ref{table:RegionIC}. We regard the examination of the emulator under this initial condition as a stringent quality test, which demonstrates the model's flexibility beyond the training range of Table~\ref{table:RegionIC} and provides insights into possible overfitting.

Figure~\ref{fig:single_cell_nocomp} presents the DeepONet predictions (dashed lines) together with the ground-truth numerical integration results (solid lines). The DeepONet model almost perfectly reproduces the highly nonlinear evolution: it captures the early chemical heating driven by $\mathrm{H}_2$ formation for $3<\log\bigl(1+\Delta t/\mathrm{sec}\bigr)<7$, as well as the late-time $\mathrm{H}_2$ cooling following further $\mathrm{H}_2$ formation at $\log\bigl(1+\Delta t/\mathrm{sec}\bigr) > 9$. Although the initial chemical abundances slightly exceed the range defined in Table~\ref{table:RegionIC}, as mentioned before, the errors remain below 30\% (except for the minor species $\mathrm{H}_2^+$, whose residual exceeds 50\% around $\log\bigl(1+\Delta t/\mathrm{sec}\bigr)\approx6.5$, where the emulated curve is less sharp than the numerical-integration result), indicating that the model can emulate the evolution even when the input slightly exceeds the range defined in Table~\ref{table:RegionIC}. We have also confirmed that the errors become smaller ($\lesssim10\%$) for the initial conditions that fall within the range of Table~\ref{table:RegionIC}. These results demonstrate that the model has successfully learned the underlying physical processes.

To assess the general accuracy of DeepONet predictions, we evaluate prediction errors using independent test datasets consisting of $5.12 \times 10^5$ samples that were not used in training or validation. The relative error $\Delta_Y$ for each prediction is defined as
\begin{equation}
\Delta_Y = \left|\frac{10^{Y_{\mathrm{pred}}} - 10^{Y_{\mathrm{true}}}}{10^{Y_{\mathrm{true}}}}\right|, \label{eq:delta_Y}
\end{equation}
where $Y$ is either $\log(T/\mathrm{K})$ or $\log(y(i))$, and the subscripts ``true" and ``pred" denote the values from numerical integration and DeepONet prediction, respectively. The cumulative distribution function (CDF) of $\log(\Delta_Y)$ is shown in Figure~\ref{fig:single_cell_CDF} for each variable and each density region. The relative errors are typically smaller than $\Delta_Y = 0.05$ and in most cases smaller than $\Delta_Y = 0.1$.

Table~\ref{table:error} summarizes the fraction of predictions whose relative errors are below 5\% and 10\% for each variable in each density region. These fractions correspond to the cumulative relative frequency evaluated at $\Delta_Y=0.05$ and $0.1$, respectively.  Except for $y(\mathrm{H}_2^+)$, more than $\sim90\%$ of the predictions have errors smaller than 10\%.  The relatively large error for $y(\mathrm{H}_2^+)$ is expected, because this species is often extremely rare and its logarithmic value is sensitive to numerical noise.  Nonetheless, because more than $\sim75\%$ of the $y(\mathrm{H}_2^+)$ predictions have errors below 10\% and $y(\mathrm{H}_2^+)$ is typically below $10^{-10}$, this deviation is unlikely to affect the overall evolution.

Temperature is a key variable for star formation because it directly affects the gas pressure and hence the gas dynamics. Our DeepONet-based emulator reproduces the temperature with very high accuracy: in more than 99\% (97\%) of the test cases, the temperature error is below 10\% (5\%), demonstrating the effectiveness of the emulator.

%%%%%%%%%%%%%%%%%%%%%%%%%%%%%%%%%%%%%%%%%%%%%%%%%%%%%%%%%%%%%%%%%%%%%%%%%%

\subsubsection{Emulating thermal and chemical evolution in a one-zone calculation}
\label{sec:euml-therm-chem}

To test whether our DeepONet-based emulator can reproduce the thermochemical evolution in Pop III star formation, we perform a one-zone collapse calculation using predictions from the emulator. In this section, we adopt a timestep of $\Delta t = 0.1t_{\mathrm{ff}}$ and update the state with simple iterative predictions, without the timescale-based modification described in Section~\ref{section:timescale-based_method}.

Figure~\ref{fig:onezone_nocomp} shows the results of the one-zone calculations. The upper and lower panels show the temperature and chemical evolution, respectively. For $n_{\mathrm{H}} \leq 10^{18}\,\mathrm{cm^{-3}}$, the solid lines show the ground-truth results obtained by numerically integrating Eqs.~\eqref{ydot} and \eqref{Tdot}, while the dashed lines show the results of DeepONet predictions. For $n_{\mathrm{H}} \ge 10^{18}\,\mathrm{cm^{-3}}$, where the chemical equilibrium is a good approximation, the solid lines show the ground-truth results obtained by numerically integrating Eq.~\eqref{Tdot} with abundances given by the Saha equation, while the dashed lines show adiabatic evolution along a sequence of chemical equilibrium states obtained by interpolating a precomputed table (see Appendix~\ref{sec:app-table-equilibrium}). 

As shown in Figure~\ref{fig:onezone_nocomp}, the DeepONet-based emulator accurately reproduces the reference evolution for $n_\mathrm{H} \le 10^{18}\,\mathrm{cm^{-3}}$, where nonequilibrium chemistry and radiative cooling are essential. In addition, the table-interpolation method also reproduces the reference evolution well in the high-density equilibrium and adiabatic regime until the gas reaches the protostellar density of $n_\mathrm{H} \sim 10^{22}\,\mathrm{cm^{-3}}$.

These results demonstrate that our emulator can follow the thermochemical evolution for $4.5\times10^{-3}\,\mathrm{cm^{-3}} \le n_\mathrm{H} \le 10^{18}\,\mathrm{cm^{-3}}$ and, when combined with the table-interpolation method, across the entire density range relevant to Pop~III star formation ($4.5\times10^{-3} \,\mathrm{cm^{-3}}\le n_\mathrm{H} \le 10^{22}\,\mathrm{cm^{-3}}$). To the best of our knowledge, this is the first study to successfully emulate thermochemical evolution using NNs across such a wide density range.

It should be noted, however, that simple DeepONet predictions can reproduce the reference evolution only when the timestep is relatively long (e.g., $\Delta t = 0.1\,t_{\mathrm{ff}}$). As shown in Section~\ref{sec:one-zone-iterative}, simply iterating the predictions with short timesteps leads to deviations from the reference evolution, thereby requiring the timescale-based method.

%%%%%%%%%%%%%%%%%%%%%%%%%%%%%%%%%%%%%%%%%%%%%%%%%%%%%%%%%%%%%%%%%%%%%%%%%%
%%%%%%%%%%%%%%%%%%%%%%%%%%%%%%%%%%%%%%%%%%%%%%%%%%%%%%%%%%%%%%%%%%%%%%%%%%

\subsection{Iterative Predictions with Short Timesteps}
\label{sec:cases-iter-updat}

Next, we address the issue of unphysical evolution caused by iterative predictions with short timesteps. We will see that this problem can be mitigated by our timescale-based method (Section~\ref{section:timescale-based_method}).

In Section~\ref{sec:learn-char-timesc}, we evaluate the performance of the DGM-based timescale estimator used in our timescale-based method. We then demonstrate the effectiveness of our method in two test cases: evolution from a fixed-density initial state (Section~\ref{sec:iter-updat-from}) and one-zone collapse calculations (Section~\ref{sec:one-zone-iterative}). In both cases, simple iteration of DeepONet predictions fails to reproduce the true evolution, but our timescale-based method yields robust predictions even under short iterative timesteps.

\subsubsection{Estimating characteristic timescales}
\label{sec:learn-char-timesc}

In addition to the DeepONet-based NNs for predictions of thermochemical evolution, we construct DGM-based NNs to estimate the timescale $\tau_\epsilon$, that is, the interval over which physical evolution changes a target variable by a fraction $\epsilon$. This timescale is used in the timescale-based method given by Eq.~\eqref{eq:timescale-based}. As mentioned in Section~\ref{section:timescale-based_method}, we adopt $\epsilon = 0.1$ when obtaining $\tau_\epsilon$, because typical errors of the DeepONet predictions are significantly smaller than 10\,\% for all variables except for $\mathrm{H}_2^{+}$, whose abundance is very small and hardly affects the overall evolution (Figure~\ref{fig:single_cell_CDF}; Table~\ref{table:error}).

To assess the accuracy of the DGM estimator of $\tau_\epsilon$, we construct a test data set of $5.12 \times 10^{5}$ samples that is independent of the training and validation sets and examine the resulting error distribution.  
The estimation error is defined as
\begin{equation}
\Delta_\tau = \left|Y_{\tau,\mathrm{pred}} - Y_{\tau,\mathrm{true}}\right|,
\label{eq:delta_chemcool}
\end{equation}
where $Y_\tau$ is a function of $\tau_\epsilon$ introduced in Eq.~\eqref{eq:Y_tau}, and the subscripts ``true" and ``pred" refer to the numerically computed ground-truth value and the NN estimate, respectively. As described in Section~\ref{section:timescale-based_method}, the network is trained to minimize a loss function based on $Y_\tau$.

Figure~\ref{fig:t_chemcool_CDF} shows the CDF of $\Delta_\tau$ for each target variable in each density region (Section~\ref{section:timescale-based_method}). Most of the errors are below $\Delta_\tau = 10^{-1}$; the only exception is temperature in Region~4, yet the errors are still concentrated below $10^{-0.5}$. The DGM-based NN estimates the characteristic timescales with reasonable accuracy.

This level of accuracy is sufficient for the timescale-based method to ensure robust iterative predictions. By construction, our method is not sensitive to the exact value of $\tau_\epsilon$, because the update formula (Eq.~\eqref{eq:timescale-based}) incorporates $\tau_\epsilon$ only indirectly. We will confirm that our timescale-based method with the DGM timescale estimator successfully models the thermochemical evolution in the following sections.

%%%%%%%%%%%%%%%%%%%%%%%%%%%%%%%%%%%%%%%%%%%%%%%%%%%%%%%%%%%%%%%%%%%%%%%%%%
%%%%%%%%%%%%%%%%%%%%%%%%%%%%%%%%%%%%%%%%%%%%%%%%%%%%%%%%%%%%%%%%%%%%%%%%%%

\subsubsection{Short-timestep iterative predictions at fixed density}
\label{sec:iter-updat-from}

In this section, we study iterative updates of the thermochemical states at fixed density, starting from a given initial condition. We show that simply iterating DeepONet predictions can lead to unphysical evolution either through error accumulation \citep{borBridging2025} or through missed long-term transitions. We also show that this issue is resolved by the timescale-based method introduced in Section~\ref{section:timescale-based_method}.

To begin with, we employ the same setup as in Section~\ref{section:error_main} but update the state with short iterative time steps. We define 1000 outer steps, $\log(1+\Delta t_0/\mathrm{sec}) = (i/1000)\log(1+\Delta t_{\mathrm{max}}/\mathrm{sec})$ ($i = 1,\dots,1000$), and divide each into 1000 equal substeps, $\Delta t=\Delta t_0^{(i)}/1000$. At each substep, we update the temperature and chemical abundances using two emulators: one simply updates the state with the DeepONet predictions, and the other uses the update formula of the timescale-based method (Eq.~\eqref{eq:timescale-based}).

Figure~\ref{fig:fixed_density_merge}(a) shows the thermochemical evolution obtained by iteratively applying the two emulators. Here, we use the same initial conditions as in Figure~\ref{fig:single_cell_nocomp}. Dashed and dotted lines show the results from simple DeepONet iterations and from the timescale-based method, respectively. Although the single-step DeepONet predictions accurately reproduce the reference evolution (Figure~\ref{fig:single_cell_nocomp}), simply repeating the DeepONet updates with short timesteps causes large deviations. The results from the timescale-based method, however, remain accurate even under repeated application.

The emulator using simple DeepONet predictions does not enforce hydrogen–nucleus conservation or charge neutrality at each substep, but we have confirmed that adding these constraints does not improve the results. Note that the timescale-based method applies both conservation laws at every substep (Section~\ref{section:timescale-based_method}).

To identify the origin of these deviations, we examine the step-by-step evolution. Figure~\ref{fig:fixed_density_merge}(b) shows the step-by-step temperature evolution in a case with $\Delta t_0=\Delta t_{\max}=4.4\times10^{9}\,\mathrm{sec}$, where simple DeepONet iterations miss long-term transitions. The horizontal axis shows the cumulative elapsed time, $\Delta t_\mathrm{elapse}=i\,\Delta t$ ($i=0,1,\dots,1000$), corresponding to the iteration count. The solid black line shows the ground-truth numerical integration; the orange dashed line with markers shows the timescale-based method; and the light-blue dotted line with markers shows simple DeepONet iterations. The markers indicate the initial state and the 1st, 10th, 100th, and 1000th iterations.

At the first substep ($\Delta t=4.4\times10^{6}\,\mathrm{sec}$), both emulators raise the temperature to the correct plateau value, $T\simeq2\times10^{3},\mathrm{K}$, because the physical change is large ($\Delta t>\tau_\epsilon$) and the timescale-based method yields the simple DeepONet value. They remain on this plateau until $\Delta t\approx10^{9}\,\mathrm{sec}$. At later times, simply iterating the DeepONet updates misses the slow $\mathrm{H}_2$ cooling and overestimates the temperature, because the relatively small physical change at each substep is masked by larger prediction errors. In contrast, the timescale-based method continues to follow the reference evolution.

Next, we examine a case in which the accumulation of prediction errors is evident. Figure~\ref{fig:fixed_density_merge}(c) is similar to panel~(a) but assumes a fixed density of $n_\mathrm{H}=10^3\,\mathrm{cm}^{-3}$ and an initial temperature of $T=300\,\mathrm{K}$. In this setup, the reference evolution remains nearly constant for all variables, which makes this setup well suited for investigating error accumulation under simple iterations. The simple iterative updates fail to keep the variables nearly constant, whereas the timescale-based method reproduces the nearly constant reference evolution with only gradual changes.

Figure~\ref{fig:fixed_density_merge}(d) shows the step-by-step temperature evolution for the case in panel~(c) with $\Delta t_0=4.0\times10^{10}\,\mathrm{sec}$. The figure is similar to panel~(b), but the bottom axis denotes the iteration count, with the top axis showing the corresponding elapsed time $\Delta t_\mathrm{elapse}$. This choice of axes makes the random-walk behavior of the simple iterative updates clearer.

The simple iterations fail to maintain a constant temperature: the temperature first drops and then rises in a random-walk manner, as expected when the accumulation of prediction errors dominates the evolution. The final temperature reaches $600\,\mathrm{K}$, twice the constant reference value of $300\,\mathrm{K}$. In contrast, the timescale-based method accurately reproduces the constant evolution.

To summarize, our timescale-based method both suppresses the accumulation of prediction errors and preserves long-term evolution, even under iterative updates with short timesteps. These results suggest that our timescale-based emulator can robustly predict the thermochemical evolution under many short timesteps. In the next section, we verify this capability in one-zone collapse calculations.

%%%%%%%%%%%%%%%%%%%%%%%%%%%%%%%%%%%%%%%%%%%%%%%%%%%%%%%%%%%%%%%%%%%%%%%%%%
%%%%%%%%%%%%%%%%%%%%%%%%%%%%%%%%%%%%%%%%%%%%%%%%%%%%%%%%%%%%%%%%%%%%%%%%%%

\subsubsection{One-zone calculations with short-timestep iterative predictions}
\label{sec:one-zone-iterative}

In this section, we evaluate the timescale-based method in one-zone collapse calculations. While Section~\ref{sec:euml-therm-chem} used a relatively large time step, $\Delta t=0.1\,t_{\mathrm{ff}}$, we adopt much smaller time steps of $\Delta t = 10^{-2}$, $10^{-3}$, and $10^{-4}\,t_{\mathrm{ff}}$ to mimic the short timesteps that can occur in hydrodynamic simulations. We compare results from the simple DeepONet emulator and the timescale-based method, focusing on issues that emerge at short timesteps.

Figure~\ref{fig:onezone_complement} shows the temperature evolution in one-zone calculations with different timesteps.  Dotted and dashed lines show the results from the simple DeepONet emulator and the timescale-based method, respectively. The solid line shows the ground-truth numerical integration result.

The simple DeepONet updates deviate increasingly from the reference evolution as $\Delta t$ decreases, with significant departures seen when $\Delta t \lesssim 10^{-2}\,t_{\mathrm{ff}}$. We omit the case with $\Delta t=10^{-4}\,t_{\mathrm{ff}}$ in the figure, as the deviation is already significant at $\Delta t=10^{-3}\,t_{\mathrm{ff}}$.

The results from the timescale-based method agree well with the reference evolution even for $\Delta t\leq10^{-2}\,t_{\mathrm{ff}}$. We show only the $10^{-3}\,t_{\mathrm{ff}}$ case, since the results for $\Delta t=10^{-1}$, $10^{-2}$, and $10^{-4}\,t_{\mathrm{ff}}$ are approximately the same.

These results confirm that our timescale-based method can emulate the thermochemical evolution in Pop III star formation accurately and robustly even under extremely small timesteps. Our approach is therefore suited for future incorporation into hydrodynamic simulations.

\begin{deluxetable*}{lccccc}
\tablewidth{0pt}
\tablecaption{Summary of computational time for single-step updates in each region.\label{table:RegionTime}}
\tablehead{& Region 0 & Region 1 & Region 2 & Region 3 & Region 4}
\startdata
    Mean [CPUms] (NI)                      & $21$ & $6.6$ & $2.7$ & $3.9$ & $3.6$\\
    Mean [CPUms] (NN)                      & $0.53$ & $0.65$ & $0.75$ & $0.74$ & $0.74$\\
    Mean speedup [$\times$] (NN vs NI)    & $40$ & $10$ & $3.7$ & $5.2$ & $4.9$\\
    Standard deviation [CPUms] (NI)        & $60$ & $11$ & $1.6$ & $1.6$ & $1.4$\\
    Standard deviation [CPUms] (NN)        & $0.11$ & $0.12$ & $0.071$ & $0.065$ & $0.059$\\
\enddata

\tablecomments{
Top and second rows: mean computational time for single-step updates for numerical integration (NI) and the DeepONet-based NN predictions (NN), respectively. Third row: speed-up of mean computational time for NN compared with NI. Fourth and fifth rows: standard deviation in the computational time for NI and NN. See the note of Table~\ref{table:error} for the definition of each density region.
}

\end{deluxetable*}

%%%%%%%%%%%%%%%%%%%%%%%%%%%%%%%%%%%%%%%%%%%%%%%%%%%%%%%%%%%%%%%%%%%%%%%%%%
%%%%%%%%%%%%%%%%%%%%%%%%%%%%%%%%%%%%%%%%%%%%%%%%%%%%%%%%%%%%%%%%%%%%%%%%%%

\section{Discussion}\label{discussion}
\subsection{Computational Time}
\label{discussion:comp-time}

To assess the speed-up provided by the NN models, we measure their computational time on an AMD EPYC 7643 CPU and an NVIDIA A6000 GPU. Section~\ref{sec:time-for-single--update} examines the cost of single-step predictions, and Section~\ref{sec:time-for-one-zone-model} discusses the computational cost when the emulator is applied to one-zone collapse calculations.

%%%%%%%%%%%%%%%%%%%%%%%%%%%%%%%%%%%%%%%%%%%%%%%%%%%%%%%%%%%%%%%%%%%%%%%%%%

\subsubsection{Single-step predictions}
\label{sec:time-for-single--update}

First, we measure the computational times for single-step DeepONet predictions and for conventional numerical integration. Both methods are tested on the same set of $5.12\times10^{5}$ random initial conditions, with a timestep of $\Delta t=\Delta t_{\max}$. Numerical integration is performed under the constraint that the relative change of any variable (temperature or any chemical abundance) at each substep does not exceed 10\% (see also Section~\ref{sec:dataset-generation}). 

Table~\ref{table:RegionTime} summarizes the mean and standard deviation of the CPU computational times for numerical integration and DeepONet prediction in each density region. Roughly speaking, DeepONet prediction is about $10$ times faster than numerical integration. Numerical integration in Region~0 takes longer than in the other regions, since $\Delta t=\Delta t_{\max}=10^{0.5}\,t_\mathrm{ff}\propto n_\mathrm{H}^{-0.5}$ (Section~\ref{sec:dataset-generation}) is larger and thus requires more substeps to satisfy the 10\% relative-change constraint. Note that these time measurements are for simple DeepONet predictions. For the timescale-based method, the cost increases by about 50\%, due to the extra cost of the DGM model for estimating $\tau_\epsilon$.

DeepONet prediction also reduces the standard deviation of computational times by more than two orders of magnitude. The large spread observed in numerical integration is due to its sensitivity to the initial state: samples far from equilibrium require many substeps for numerical integration. Such variability limits parallel efficiency because faster cells must wait for slower ones.

In addition, we measure the GPU computational times for DeepONet prediction, combining all $5.12\times10^{5}$ input samples into a single batch processed at once. On the GPU, DeepONet prediction is more than three (two) orders of magnitude faster than numerical integration (DeepONet predictions on the CPU). In the measurement, we use an NVIDIA RTX A6000 with 10,752 CUDA cores. We also test smaller batch sizes and find that the computational time does not change significantly until the batch size becomes smaller than 5,000, suggesting that the GPU becomes saturated around a batch size of $\sim$5,000.

These results suggest that NN emulators, such as our DeepONet model, are well suited to large-scale parallel simulations owing to their low variance in computational times. Furthermore, such emulators have the potential to enable even faster modeling of thermochemical evolution in GPU-accelerated simulations.

%%%%%%%%%%%%%%%%%%%%%%%%%%%%%%%%%%%%%%%%%%%%%%%%%%%%%%%%%%%%%%%%%%%%%%%%%%

\subsubsection{One-zone calculations}
\label{sec:time-for-one-zone-model}
Next, we measure the computational times of one-zone calculations with $\Delta t = 0.1\,t_{\mathrm{ff}}$ for the emulator using the timescale-based method and for conventional numerical integration. We replicate the same initial state into a batch of $32^{3}$ cells to mimic a workload in three-dimensional simulations. We measure the emulator computational time on both the CPU and the GPU. The GPU processes the entire batch in a single call, as described in Section~\ref{sec:time-for-single--update}.

On the CPU, the emulator is slightly slower than numerical integration because numerical integration requires fewer substeps under near-equilibrium one-zone conditions than for random initial conditions. In full three-dimensional simulations, however, shocks and other dynamical processes likely drive the gas far from equilibrium and increase the integration cost. Under those conditions, we expect the emulator, which is insensitive to initial conditions, to outperform numerical integration (see Section~\ref{sec:time-for-single--update}).

The GPU greatly accelerates the emulator: GPU emulation is roughly $35$ times faster than numerical integration. Given the growing availability of GPUs and the recent development of GPU-supported hydrodynamic codes, such as \texttt{CHOLLA} \citep{schneiderCHOLLA2015a},  \texttt{GAMER-2} \citep{schiveGAMER22018}, \texttt{QUOKKA} \citep{Wibking:2022}, and \texttt{AthenaK} \citep{stoneAthenaK2024}, we expect that NN-based emulators for thermochemical evolution, including ours, will provide substantial benefits for future hydrodynamic simulations.

%%%%%%%%%%%%%%%%%%%%%%%%%%%%%%%%%%%%%%%%%%%%%%%%%%%%%%%%%%%%%%%%%%%%%%%%%%
%%%%%%%%%%%%%%%%%%%%%%%%%%%%%%%%%%%%%%%%%%%%%%%%%%%%%%%%%%%%%%%%%%%%%%%%%%

\subsection{Comparison with Previous Work}
We first compare our results with \citet{brancaEmulating2024}, who used DeepONet to model the low-density ISM ($n_\mathrm{H} \lesssim10^{4}\,\mathrm{cm}^{-3}$) under an external radiation field. In contrast, we have considered a much broader density range, $10^{-3}\,\mathrm{cm}^{-3}\le n_{\mathrm{H}}\le10^{18}\,\mathrm{cm}^{-3}$, relevant to Pop III star formation, and followed the evolution without radiation. Moreover, whereas \citet{brancaEmulating2024} trained their NNs for short timesteps of $\Delta t$ from 0 to 1 kyr using a linear time scale, we trained ours for the much wider timesteps of $\Delta t$ from 0 sec to $10^{0.5}\,t_\mathrm{ff}$ (which is $\gtrsim1$ Myr at $n_{\mathrm{H}}\lesssim10^{4}\,\mathrm{cm}^{-3}$) using a logarithmic time scaling, as in \citet{brancaNeural2023}.

We obtained overall accuracies comparable to those of \citet{brancaEmulating2024}, except for $\mathrm{H}^-$ and $\mathrm{H}_2^+$, whose errors are about an order of magnitude larger than those reported in the literature. This reduced accuracy is not critical, since these species are often rare and have a negligible impact on the global evolution. The validity of our NN-based emulators has been demonstrated by their successful application to one-zone calculations (Sections~\ref{sec:euml-therm-chem} and \ref{sec:one-zone-iterative}).

Next, we discuss the issue of simple iterative NN updates leading to unphysical evolution, in part due to the accumulation of prediction errors \citep{borBridging2025}. For iterative updates at fixed density, we confirmed this behavior and showed that such updates can also fail to capture long-term changes (Section~\ref{sec:iter-updat-from}). Similarly, we found that simple iterative NN updates lead to significant deviations from the reference thermal trajectory in one-zone calculations with short timesteps (Section~\ref{sec:one-zone-iterative}).

The timescale-based method introduced in Section~\ref{section:timescale-based_method} resolves these issues (Sections~\ref{sec:iter-updat-from} and \ref{sec:one-zone-iterative}). This method enables robust iterative predictions even under short-timestep updates, as typically required in hydrodynamic simulations.

Our timescale-based method differs from previous approaches proposed to mitigate issues arising from iterative NN updates in autoencoder-based models, such as injecting Gaussian noise during training \citep{holdshipChemulator2021} and training the iterative updates themselves \citep{maesMACE2024}. These methods are known to increase one-step prediction errors \citep{holdshipChemulator2021,maesMACE2024}. In contrast, our method keeps one-step prediction errors low because the original NN prediction is used when the predicted change exceeds the estimated prediction error.

%%%%%%%%%%%%%%%%%%%%%%%%%%%%%%%%%%%%%%%%%%%%%%%%%%%%%%%%%%%%%%%%%%%%%%%%%%
%%%%%%%%%%%%%%%%%%%%%%%%%%%%%%%%%%%%%%%%%%%%%%%%%%%%%%%%%%%%%%%%%%%%%%%%%%

\section{Conclusion}
\label{sec:conclusion}

In this study, we developed a neural–network (NN) emulator that models the thermal and chemical evolution of Pop III star formation over the wide density range $10^{-3}\,\mathrm{cm}^{-3}\le n_{\mathrm H}\le10^{18}\,\mathrm{cm}^{-3}$. To make accurate predictions over such a wide range, we divided the range into five subregions and trained separate DeepONet models for each region.

Furthermore, to ensure robust predictions even when many short-timestep iterations are required, we introduced a new update method based on the timescale for significant variation. Specifically, we constructed DGM-based NNs to estimate the timescale $\tau_\epsilon$ defined as the minimum time over which a physical change significantly exceeds the typical errors of DeepONet predictions. If the given timestep $\Delta t$ is shorter than $\tau_\epsilon$, the change over $\Delta t$ is computed by scaling the DeepONet change over $\tau_\epsilon$ by $\Delta t/\tau_\epsilon$.

We demonstrated that our emulator accurately reproduces the thermochemical evolution. It performs well both for fixed-density evolution and for one-zone calculations that mimic typical star-formation conditions. For $5.12\times10^{5}$ random states, the emulator matches gas temperature and the abundances to $<10\,\%$ relative error in $\gtrsim90\,\%$ of cases, except for slightly larger scatter for H$_2^{+}$. Even when the timestep is greatly reduced and many iterations are needed, the newly developed timescale-based method robustly reproduces the thermochemical evolution. In terms of performance, our tests indicated that the emulator is about ten times faster than conventional numerical integration on the CPU, and more than a thousand times faster in batched GPU calculations.

NN emulators of thermochemical evolution are well suited for integration into three-dimensional star-formation simulations. The computational speedup and the better parallel efficiency are particularly advantageous for large-scale parallel or GPU-accelerated calculations. Although our current models target a metal-free and radiation-free gas, we plan to incorporate additional physics, such as metals and radiative feedback, to broaden the applicability. We also plan to integrate our emulator into hydrodynamic simulations of Pop III star formation, opening a new path to exploring star formation using simulations accelerated by chemical emulators.

\begin{acknowledgments}
The authors thank Hajime Fukushima, Kazutaka Kimura, Kazuyuki Omukai, Keiichi Maeda, Takashi Hosokawa, and Takashi Okamoto for fruitful discussions and comments. The numerical calculations were performed on the GPU system and HPE Cray XD2000 at the Center for Computational Astrophysics at the National Astronomical Observatory of Japan. This work was supported by JSPS KAKENHI Grant Numbers 24H00002 (KS). This research was also supported by MEXT as ``Program for Promoting Researches on the Supercomputer Fugaku'' (Structure and Evolution of the Universe Unraveled by Fusion of Simulation and AI, JPMXP1020230406).
\end{acknowledgments}

%%%%%%%%%%%%%%%%%%%%%%%%%%%%%%%%%%%%%%%%%%%%%%%%%%%%%%%%%%%%%%%%%%%%%%%%%%
%%%%%%%%%%%%%%%%%%%%%%%%%%%%%%%%%%%%%%%%%%%%%%%%%%%%%%%%%%%%%%%%%%%%%%%%%%

\appendix

\section{Chemical Reactions}
\label{sec:app-chemical-reactions}

In this study, we considered a total of 18 chemical reactions relevant to primordial gas chemistry. These include key collisional and associative processes that govern the formation and destruction of molecular hydrogen and other species in metal-free environments. The full list of reactions is provided in Table~\ref{table:chemical_reaction}.

\section{Table-interpolation Method for High-density Equilibrium and Adiabatic Regime}
\label{sec:app-table-equilibrium}
For $n_{\mathrm{H}} \geq 10^{18}\,\mathrm{cm^{-3}}$, reaction timescales are so short that the chemical evolution proceeds along a sequence of chemical equilibrium states set by density and temperature according to the Saha equation \citep{sahaLIII1920}. Moreover, at such high densities, radiative cooling via line and continuum emission is so inefficient that the specific entropy is approximately conserved and the gas can be treated as adiabatic \citep{omukaiPrimordial2001}. In the operator-splitting approach, the total energy, consisting of internal and chemical energy, does not change during the stage of chemical and thermal updates given by Eqs.~\eqref{ydot} and \eqref{Tdot}.

Therefore, in this high-density equilibrium and adiabatic regime, the updated chemical abundances $y(i)$ and the temperature $T$ can be uniquely determined from the pair of variables $n_{\mathrm{H}}$ and the total energy per hydrogen atom $E_\mathrm{tot}$. Here, $E_\mathrm{tot}$ is given by
\begin{align}
    E_\mathrm{tot} &= U + E_\mathrm{chem}\,,
\end{align}
where $U$ and $E_\mathrm{chem}$ are the internal energy and chemical energy per hydrogen atom, respectively. We calculate the internal energy $U$ as
\begin{align}
U = \frac{3}{2}k_\mathrm{B} T
    \left(\sum_{i=1}^6 y(i) + y_\mathrm{He}\right)
    + \left(\int_{0\,\mathrm{K}}^T c_{v,\mathrm{H}_2}(T')\,\mathrm{d}T' - \frac{3}{2}k_\mathrm{B} T \right) y(\mathrm{H}_2)\,,
\end{align}
with the temperature-dependent heat capacity of molecular hydrogen $c_{v,\mathrm{H}_2}(T)$ (\citealt{Sugimura:2023}; see \citealt{black1975Evolution} for the calculation method), which accounts for rovibrational excitations. The chemical energy $E_\mathrm{chem}$ is given by
\begin{align}
E_\mathrm{chem} = 
    \sum_{i=1}^6 y(i)\,\chi(i)\,,
\end{align}
where $\chi(i)$ is the chemical energy of species $i$ measured with respect to $\mathrm{H}_2$. We set $\chi(\mathrm{H})=216.034$, $\chi(\mathrm{H}_2)=0$, $\chi(\mathrm{e}^-)=0$, $\chi(\mathrm{H}^+)=1528.084$, $\chi(\mathrm{H}^-)=143.246$, and $\chi(\mathrm{H}_2^+)=1488.365$ (in $\mathrm{kJ\,mol^{-1}}$), from the KIDA database \citep{wakelamKINETIC2012}. 

Given that the updated $y(i)$ and $T$ are determined by $n_{\mathrm{H}}$ and $E_\mathrm{tot}$, the chemical and thermal updates obtained by numerically integrating Eqs.~\eqref{ydot} and \eqref{Tdot} can be replaced by interpolating a precomputed two-dimensional table on a grid in $(n_{\mathrm{H}},\, E_\mathrm{tot})$ space. Because the table is two-dimensional, the table size needed for accurate interpolation is reasonably small. Therefore, we adopt the table-interpolation method rather than NN-based emulators in this high-density regime. Like NN-based emulators, the table-interpolation method yields a small standard deviation of computational time and enables efficient batch processing, making it well suited to large-scale parallel and GPU-accelerated simulations.

In our implementation, we tabulate $n_{\mathrm{H}}$ from $10^{18}$ to $10^{22}\,\mathrm{cm^{-3}}$ on a 1000-point logarithmic grid. We also tabulate $E_{\mathrm{tot}}$ on a 1000-point logarithmic grid over the range corresponding to $10\,\mathrm{K} < T < 10^5\,\mathrm{K}$, which covers the values relevant for Pop~III star formation. To construct the table, for each grid point $(n_{\mathrm{H}}, E_\mathrm{tot})$, we obtain the combination of $y(i)$ and $T$ that satisfies the Saha equation with the total energy equal to the given $E_\mathrm{tot}$. To compute the updated $y(i)$ and $T$, we perform linear interpolation of this table in logarithmic space. We demonstrate the accuracy of this table-interpolation method by applying it to a one-zone collapse calculation in Section~\ref{sec:euml-therm-chem}.

\begin{deluxetable*}{rlcrlc}
\tablewidth{0pt}
\tablecaption{Chemical reactions considered in this study\label{table:chemical_reaction}}
\tablehead{Index & Reactions & References & Index & Reactions & References}
\startdata
    1 & $\mathrm{H} + \mathrm{e}^- \rightarrow \mathrm{H}^+ + 2\mathrm{e}^-$  & [1] & 10 & $\mathrm{H} + \mathrm{e}^- \rightarrow \mathrm{H}^- + \gamma$           & [4]  \\
    2 & $\mathrm{H}^+ + \mathrm{e}^- \rightarrow \mathrm{H} + \gamma$         & [2] & 11 & $2\mathrm{H} \rightarrow \mathrm{H}^+ + \mathrm{e}^- + \mathrm{H}$      & [7] \\
    3 & $\mathrm{H}^- + \mathrm{H} \rightarrow \mathrm{H}_2 + \mathrm{e}^-$   & [3] & 12 & $\mathrm{H}^- + \mathrm{e}^- \rightarrow \mathrm{H} + 2\mathrm{e}^-$    & [1]    \\
    4 & $\mathrm{H}_2 + \mathrm{H}^+ \rightarrow \mathrm{H}_2^+ + \mathrm{H}$ & [4] & 13 & $\mathrm{H}^- + \mathrm{H}^+ \rightarrow \mathrm{H}_2^+ + \mathrm{e}^-$ & [4]  \\
    5 & $\mathrm{H}_2 + \mathrm{e}^- \rightarrow 2\mathrm{H} + \mathrm{e}^-$  & [4] & 14 & $\mathrm{H}^- + \mathrm{H}^+ \rightarrow 2\mathrm{H}$                   & [4]  \\
    6 & $\mathrm{H}_2 + \mathrm{H} \rightarrow 3\mathrm{H}$                   & [5] & 15 & $\mathrm{H} + \mathrm{H}^+ \rightarrow \mathrm{H}_2^+ + \gamma$         & [4]  \\
    7 & $3\mathrm{H} \rightarrow \mathrm{H}_2 + \mathrm{H}$                   & [6] & 16 & $\mathrm{H}_2^+ + \mathrm{H} \rightarrow \mathrm{H}_2 + \mathrm{H}^+$   & [4]  \\
    8 & $2\mathrm{H} + \mathrm{H}_2 \rightarrow 2\mathrm{H}_2$                & [7] & 17 & $\mathrm{H}_2^+ + \mathrm{e}^- \rightarrow 2\mathrm{H}$                 & [4]  \\
    9 & $2\mathrm{H}_2 \rightarrow 2\mathrm{H} + \mathrm{H}_2$                & [7,8] & 18 & $\mathrm{H}_2^+ + \mathrm{H}^- \rightarrow \mathrm{H}_2 + \mathrm{H}$   & [9]
\enddata
\tablerefs{[1] \citet{abelModeling1997}, [2] \citet{glover2007star}, [3] \citet{kreckelExperimental2010}, [4] \citet{galliChemistry1998}, [5] \citet{martinMaster1996}, [6] \citet{forreyRATE2013}, [7] \citet{pallaPrimordial1983}, [8] \citet{shapiro1987hydrogen}, [9] \citet{millar1991gas}}

\end{deluxetable*}

%%%%%%%%%%%%%%%%%%%%%%%%%%%%%%%%%%%%%%%%%%%%%%%%%%%%%%%%%%%%%%%%%%%%%%%%%%
%%%%%%%%%%%%%%%%%%%%%%%%%%%%%%%%%%%%%%%%%%%%%%%%%%%%%%%%%%%%%%%%%%%%%%%%%%


\begin{thebibliography}{}
\expandafter\ifx\csname natexlab\endcsname\relax\def\natexlab#1{#1}\fi
\providecommand{\url}[1]{\href{#1}{#1}}
\providecommand{\dodoi}[1]{doi:~\href{http://doi.org/#1}{\nolinkurl{#1}}}
\providecommand{\doeprint}[1]{\href{http://ascl.net/#1}{\nolinkurl{http://ascl.net/#1}}}
\providecommand{\doarXiv}[1]{\href{https://arxiv.org/abs/#1}{\nolinkurl{https://arxiv.org/abs/#1}}}

% type= misc
\bibitem[{M. Abadi {et~al.}(2015)Abadi, Agarwal, Barham, Brevdo, Chen, Citro, Corrado, Davis, Dean, Devin, Ghemawat, Goodfellow, Harp, Irving, Isard, Jia, Jozefowicz, Kaiser, Kudlur, Levenberg, Man\'{e}, Monga, Moore, Murray, Olah, Schuster, Shlens, Steiner, Sutskever, Talwar, Tucker, Vanhoucke, Vasudevan, Vi\'{e}gas, Vinyals, Warden, Wattenberg, Wicke, Yu, \& Zheng}]{tensorflow2015-whitepaper}
Abadi, M., Agarwal, A., Barham, P., {et~al.} 2015, {TensorFlow}: Large-Scale Machine Learning on Heterogeneous Systems, \url{https://www.tensorflow.org/}

% type= article
\bibitem[{T. Abel {et~al.}(1997)Abel, Anninos, Zhang, \& Norman}]{abelModeling1997}
Abel, T., Anninos, P., Zhang, Y., \& Norman, M.~L. 1997, \bibinfo{title}{Modeling Primordial Gas in Numerical Cosmology,} \na, 2, 181, \dodoi{10.1016/S1384-1076(97)00010-9}

% type= article
\bibitem[{T. {Abel} {et~al.}(2002){Abel}, {Bryan}, \& {Norman}}]{Abel:2002}
{Abel}, T., {Bryan}, G.~L., \& {Norman}, M.~L. 2002, \bibinfo{title}{{The Formation of the First Star in the Universe},} Science, 295, 93, \dodoi{10.1126/science.1063991}

% type= article
\bibitem[{D.~C. Black \& P. Bodenheimer(1975)Black \& Bodenheimer}]{black1975Evolution}
Black, D.~C., \& Bodenheimer, P. 1975, \bibinfo{title}{Evolution of Rotating Interstellar Clouds. {{I}}. {{Numerical}} Techniques.,} \apj, 199, 619, \dodoi{10.1086/153729}

% type= article
\bibitem[{L. Branca \& A. Pallottini(2023)Branca \& Pallottini}]{brancaNeural2023}
Branca, L., \& Pallottini, A. 2023, \bibinfo{title}{Neural Networks: Solving the Chemistry of the Interstellar Medium,} \mnras, 518, 5718, \dodoi{10.1093/mnras/stac3512}

% type= article
\bibitem[{L. Branca \& A. Pallottini(2024)Branca \& Pallottini}]{brancaEmulating2024}
Branca, L., \& Pallottini, A. 2024, \bibinfo{title}{Emulating the Interstellar Medium Chemistry with Neural Operators,} \aap, 684, A203, \dodoi{10.1051/0004-6361/202449193}

% type= article
\bibitem[{V. {Bromm} {et~al.}(2002){Bromm}, {Coppi}, \& {Larson}}]{Bromm:2002}
{Bromm}, V., {Coppi}, P.~S., \& {Larson}, R.~B. 2002, \bibinfo{title}{{The Formation of the First Stars. I. The Primordial Star-forming Cloud},} \apj, 564, 23, \dodoi{10.1086/323947}

% type= article
\bibitem[{D. {de Mijolla} {et~al.}(2019){de Mijolla}, {Viti}, {Holdship}, {Manolopoulou}, \& {Yates}}]{deMijolla:2019}
{de Mijolla}, D., {Viti}, S., {Holdship}, J., {Manolopoulou}, I., \& {Yates}, J. 2019, \bibinfo{title}{{Incorporating astrochemistry into molecular line modelling via emulation},} \aap, 630, A117, \dodoi{10.1051/0004-6361/201935973}

% type= article
\bibitem[{R.~C. Forrey(2013)Forrey}]{forreyRATE2013}
Forrey, R.~C. 2013, \bibinfo{title}{{{RATE OF FORMATION OF HYDROGEN MOLECULES BY THREE-BODY RECOMBINATION DURING PRIMORDIAL STAR FORMATION}},} \apjl, 773, L25, \dodoi{10.1088/2041-8205/773/2/L25}

% type= article
\bibitem[{H. {Fukushima} {et~al.}(2018){Fukushima}, {Omukai}, \& {Hosokawa}}]{Fukushima:2018}
{Fukushima}, H., {Omukai}, K., \& {Hosokawa}, T. 2018, \bibinfo{title}{{Upper stellar mass limit by radiative feedback at low-metallicities: metallicity and accretion rate dependence},} \mnras, 473, 4754, \dodoi{10.1093/mnras/stx2620}

% type= article
\bibitem[{D. Galli \& F. Palla(1998)Galli \& Palla}]{galliChemistry1998}
Galli, D., \& Palla, F. 1998, \bibinfo{title}{The Chemistry of the Early {{Universe}},} \aap, 335, 403

% type= inproceedings
\bibitem[{X. Glorot {et~al.}(2011)Glorot, Bordes, \& Bengio}]{glorot2011deep}
Glorot, X., Bordes, A., \& Bengio, Y. 2011, \bibinfo{title}{Deep sparse rectifier neural networks,} in Proceedings of the fourteenth international conference on artificial intelligence and statistics, JMLR Workshop and Conference Proceedings, 315--323

% type= article
\bibitem[{S.~C.~O. {Glover}(2015){Glover}}]{Glover:2015}
{Glover}, S. C.~O. 2015, \bibinfo{title}{{Simulating the formation of massive seed black holes in the early Universe - I. An improved chemical model},} \mnras, 451, 2082, \dodoi{10.1093/mnras/stv1059}

% type= article
\bibitem[{S.~C.~O. Glover \& A.-K. Jappsen(2007)Glover \& Jappsen}]{glover2007star}
Glover, S. C.~O., \& Jappsen, A.-K. 2007, \bibinfo{title}{Star Formation at Very Low Metallicity. I. Chemistry and Cooling at Low Densities,} The Astrophysical Journal, 666, 1, \dodoi{10.1086/519445}

% type= article
\bibitem[{T. {Grassi} {et~al.}(2011){Grassi}, {Merlin}, {Piovan}, {Buonomo}, \& {Chiosi}}]{Grassi:2011}
{Grassi}, T., {Merlin}, E., {Piovan}, L., {Buonomo}, U., \& {Chiosi}, C. 2011, \bibinfo{title}{{MaNN: Multiple Artificial Neural Networks for modelling the Interstellar Medium},} arXiv e-prints, arXiv:1103.0509, \dodoi{10.48550/arXiv.1103.0509}

% type= article
\bibitem[{T. Grassi {et~al.}(2022)Grassi, Nauman, Ramsey, Bovino, Picogna, \& Ercolano}]{grassiReducing2022}
Grassi, T., Nauman, F., Ramsey, J.~P., {et~al.} 2022, \bibinfo{title}{Reducing the Complexity of Chemical Networks via Interpretable Autoencoders,} \aap, 668, A139, \dodoi{10.1051/0004-6361/202039956}

% type= article
\bibitem[{A. Grichener {et~al.}(2025)Grichener, Renzo, Kerzendorf, Farmer, De~Mink, Bellinger, Chan, Chen, Farag, \& Justham}]{grichenerNuclear2025}
Grichener, A., Renzo, M., Kerzendorf, W.~E., {et~al.} 2025, \bibinfo{title}{Nuclear {{Neural Networks}}: {{Emulating Late Burning Stages}} in {{Core-collapse Supernova Progenitors}},} \apjs, 279, 49, \dodoi{10.3847/1538-4365/ade717}

% type= article
\bibitem[{J. {Heyl} {et~al.}(2023){Heyl}, {Viti}, \& {Vermari{\"e}n}}]{Heyl:2023}
{Heyl}, J., {Viti}, S., \& {Vermari{\"e}n}, G. 2023, \bibinfo{title}{{A statistical and machine learning approach to the study of astrochemistry},} Faraday Discussions, 245, 569, \dodoi{10.1039/D3FD00008G}

% type= article
\bibitem[{A.~E. Hoerl \& R.~W. Kennard(1970{\natexlab{a}})Hoerl \& Kennard}]{hoerl1970ridge1}
Hoerl, A.~E., \& Kennard, R.~W. 1970{\natexlab{a}}, \bibinfo{title}{Ridge regression: Biased estimation for nonorthogonal problems,} Technometrics, 12, 55

% type= article
\bibitem[{A.~E. Hoerl \& R.~W. Kennard(1970{\natexlab{b}})Hoerl \& Kennard}]{hoerl1970ridge2}
Hoerl, A.~E., \& Kennard, R.~W. 1970{\natexlab{b}}, \bibinfo{title}{Ridge regression: applications to nonorthogonal problems,} Technometrics, 12, 69

% type= article
\bibitem[{J. Holdship {et~al.}(2021)Holdship, Viti, Haworth, \& Ilee}]{holdshipChemulator2021}
Holdship, J., Viti, S., Haworth, T.~J., \& Ilee, J.~D. 2021, \bibinfo{title}{Chemulator: {{Fast}}, Accurate Thermochemistry for Dynamical Models through Emulation,} \aap, 653, A76, \dodoi{10.1051/0004-6361/202140357}

% type= article
\bibitem[{T. Hosokawa {et~al.}(2011)Hosokawa, Omukai, Yoshida, \& Yorke}]{hosokawaProtostellar2011}
Hosokawa, T., Omukai, K., Yoshida, N., \& Yorke, H.~W. 2011, \bibinfo{title}{Protostellar {{Feedback Halts}} the {{Growth}} of the {{First Stars}} in the {{Universe}},} Science, 334, 1250, \dodoi{10.1126/science.1207433}

% type= article
\bibitem[{D.~P. Kingma \& J. Ba(2014)Kingma \& Ba}]{kingmaAdam2014}
Kingma, D.~P., \& Ba, J. 2014, \bibinfo{title}{Adam: {{A Method}} for {{Stochastic Optimization}},} arXiv e-prints, arXiv:1412.6980, \dodoi{10.48550/arXiv.1412.6980}

% type= article
\bibitem[{R.~S. {Klessen} \& S.~C.~O. {Glover}(2023){Klessen} \& {Glover}}]{Klessen:2023}
{Klessen}, R.~S., \& {Glover}, S. C.~O. 2023, \bibinfo{title}{{The First Stars: Formation, Properties, and Impact},} \araa, 61, 65, \dodoi{10.1146/annurev-astro-071221-053453}

% type= article
\bibitem[{H. Kreckel {et~al.}(2010)Kreckel, Bruhns, {\v C}{\'i}{\v z}ek, Glover, Miller, Urbain, \& Savin}]{kreckelExperimental2010}
Kreckel, H., Bruhns, H., {\v C}{\'i}{\v z}ek, M., {et~al.} 2010, \bibinfo{title}{Experimental {{Results}} for {{H2 Formation}} from {{H}}- and {{H}} and {{Implications}} for {{First Star Formation}},} Science, 329, 69, \dodoi{10.1126/science.1187191}

% type= article
\bibitem[{L. Lu {et~al.}(2021{\natexlab{a}})Lu, Jin, Pang, Zhang, \& Karniadakis}]{luLearning2021}
Lu, L., Jin, P., Pang, G., Zhang, Z., \& Karniadakis, G.~E. 2021{\natexlab{a}}, \bibinfo{title}{Learning Nonlinear Operators via {{DeepONet}} Based on the Universal Approximation Theorem of Operators,} NatMI, 3, 218, \dodoi{10.1038/s42256-021-00302-5}

% type= article
\bibitem[{L. Lu {et~al.}(2021{\natexlab{b}})Lu, Meng, Mao, \& Karniadakis}]{luDeepXDE2021}
Lu, L., Meng, X., Mao, Z., \& Karniadakis, G.~E. 2021{\natexlab{b}}, \bibinfo{title}{{{DeepXDE}}: {{A Deep Learning Library}} for {{Solving Differential Equations}},} SIAMR, 63, 208, \dodoi{10.1137/19M1274067}

% type= article
\bibitem[{S. Maes {et~al.}(2024)Maes, De~Ceuster, {Van de Sande}, \& Decin}]{maesMACE2024}
Maes, S., De~Ceuster, F., {Van de Sande}, M., \& Decin, L. 2024, \bibinfo{title}{{{MACE}}: {{A Machine-learning Approach}} to {{Chemistry Emulation}},} \apj, 969, 79, \dodoi{10.3847/1538-4357/ad47a1}

% type= article
\bibitem[{P.~G. Martin {et~al.}(1996)Martin, Schwarz, \& Mandy}]{martinMaster1996}
Martin, P.~G., Schwarz, D.~H., \& Mandy, M.~E. 1996, \bibinfo{title}{Master {{Equation Studies}} of the {{Collisional Excitation}} and {{Dissociation}} of {{H2 Molecules}} by {{H Atoms}},} \apj, 461, 265, \dodoi{10.1086/177053}

% type= article
\bibitem[{R. {Matsukoba} {et~al.}(2019){Matsukoba}, {Takahashi}, {Sugimura}, \& {Omukai}}]{Matsukoba:2019}
{Matsukoba}, R., {Takahashi}, S.~Z., {Sugimura}, K., \& {Omukai}, K. 2019, \bibinfo{title}{{Gravitational stability and fragmentation condition for discs around accreting supermassive stars},} \mnras, 484, 2605, \dodoi{10.1093/mnras/sty3522}

% type= article
\bibitem[{T. Millar {et~al.}(1991)Millar, Bennett, Rawlings, Brown, \& Charnley}]{millar1991gas}
Millar, T., Bennett, A., Rawlings, J., Brown, P., \& Charnley, S. 1991, \bibinfo{title}{Gas phase reactions and rate coefficients for use in astrochemistry-The UMIST ratefile,} \aaps, 87, 585

% type= article
\bibitem[{T. {Nagakura} \& K. {Omukai}(2005){Nagakura} \& {Omukai}}]{Nagakura:2005}
{Nagakura}, T., \& {Omukai}, K. 2005, \bibinfo{title}{{Formation of Population III stars in fossil HII regions: significance of HD},} \mnras, 364, 1378, \dodoi{10.1111/j.1365-2966.2005.09685.x}

% type= article
\bibitem[{D. {Nakauchi} {et~al.}(2019){Nakauchi}, {Omukai}, \& {Susa}}]{Nakauchi:2019}
{Nakauchi}, D., {Omukai}, K., \& {Susa}, H. 2019, \bibinfo{title}{{Ionization degree and magnetic diffusivity in the primordial star-forming clouds},} \mnras, 488, 1846, \dodoi{10.1093/mnras/stz1799}

% type= article
\bibitem[{L.~M. Navarro {et~al.}(2023)Navarro, {Martin-Moreno}, \& Rodrigo}]{navarroSolving2023}
Navarro, L.~M., {Martin-Moreno}, L., \& Rodrigo, S.~G. 2023, \bibinfo{title}{Solving Differential Equations with Deep Learning: A Beginner's Guide,} EJPh, 45, 015803, \dodoi{10.1088/1361-6404/ad0a9f}

% type= article
\bibitem[{K. Omukai(2001)Omukai}]{omukaiPrimordial2001}
Omukai, K. 2001, \bibinfo{title}{Primordial {{Star Formation}} under {{Far-UltravioletRadiation}},} \apj, 546, 635, \dodoi{10.1086/318296}

% type= article
\bibitem[{K. Omukai \& R. Nishi(1998)Omukai \& Nishi}]{omukaiFormation1998}
Omukai, K., \& Nishi, R. 1998, \bibinfo{title}{Formation of {{Primordial Protostars}},} \apj, 508, 141, \dodoi{10.1086/306395}

% type= article
\bibitem[{F. Palla {et~al.}(1983)Palla, Salpeter, \& Stahler}]{pallaPrimordial1983}
Palla, F., Salpeter, E.~E., \& Stahler, S.~W. 1983, \bibinfo{title}{Primordial Star Formation - {{The}} Role of Molecular Hydrogen,} \apj, 271, 632, \dodoi{10.1086/161231}

% type= article
\bibitem[{P. {Palud} {et~al.}(2023){Palud}, {Einig}, {Le Petit}, {Bron}, {Chainais}, {Chanussot}, {Pety}, {Thouvenin}, {Languignon}, {Be{\v{s}}li{\'c}}, {Santa-Maria}, {Orkisz}, {S{\'e}gal}, {Zakardjian}, {Bardeau}, {Gerin}, {Goicoechea}, {Gratier}, {Guzman}, {Hughes}, {Levrier}, {Liszt}, {Le Bourlot}, {Roueff}, \& {Sievers}}]{Palud:2023}
{Palud}, P., {Einig}, L., {Le Petit}, F., {et~al.} 2023, \bibinfo{title}{{Neural network-based emulation of interstellar medium models},} \aap, 678, A198, \dodoi{10.1051/0004-6361/202347074}

% type= article
\bibitem[{A. Paszke {et~al.}(2019)Paszke, Gross, Massa, Lerer, Bradbury, Chanan, Killeen, Lin, Gimelshein, Antiga, Desmaison, K{\"o}pf, Yang, DeVito, Raison, Tejani, Chilamkurthy, Steiner, Fang, Bai, \& Chintala}]{paszkePyTorch2019}
Paszke, A., Gross, S., Massa, F., {et~al.} 2019, \bibinfo{title}{{{PyTorch}}: {{An Imperative Style}}, {{High-Performance Deep Learning Library}},} arXiv e-prints, arXiv:1912.01703, \dodoi{10.48550/arXiv.1912.01703}

% type= article
\bibitem[{M. Raissi {et~al.}(2019)Raissi, Perdikaris, \& Karniadakis}]{raissiPhysicsinformed2019}
Raissi, M., Perdikaris, P., \& Karniadakis, G. 2019, \bibinfo{title}{Physics-Informed Neural Networks: {{A}} Deep Learning Framework for Solving Forward and Inverse Problems Involving Nonlinear Partial Differential Equations,} JCoPh, 378, 686, \dodoi{10.1016/j.jcp.2018.10.045}

% type= article
\bibitem[{K.~E. {Sadanari} {et~al.}(2021){Sadanari}, {Omukai}, {Sugimura}, {Matsumoto}, \& {Tomida}}]{Sadanari:2021}
{Sadanari}, K.~E., {Omukai}, K., {Sugimura}, K., {Matsumoto}, T., \& {Tomida}, K. 2021, \bibinfo{title}{{Magnetohydrodynamic effect on first star formation: pre-stellar core collapse and protostar formation},} \mnras, 505, 4197, \dodoi{10.1093/mnras/stab1330}

% type= article
\bibitem[{K.~E. {Sadanari} {et~al.}(2023){Sadanari}, {Omukai}, {Sugimura}, {Matsumoto}, \& {Tomida}}]{Sadanari:2023}
{Sadanari}, K.~E., {Omukai}, K., {Sugimura}, K., {Matsumoto}, T., \& {Tomida}, K. 2023, \bibinfo{title}{{Non-ideal magnetohydrodynamic simulations of the first star formation: the effect of ambipolar diffusion},} \mnras, 519, 3076, \dodoi{10.1093/mnras/stac3724}

% type= article
\bibitem[{M.~N. Saha(1920)Saha}]{sahaLIII1920}
Saha, M.~N. 1920, \bibinfo{title}{{{LIII}}. {{Ionization}} in the Solar Chromosphere,} PMag, 40, 472, \dodoi{10.1080/14786441008636148}

% type= article
\bibitem[{H.-Y. Schive {et~al.}(2018)Schive, ZuHone, Goldbaum, Turk, Gaspari, \& Cheng}]{schiveGAMER22018}
Schive, H.-Y., ZuHone, J.~A., Goldbaum, N.~J., {et~al.} 2018, \bibinfo{title}{{{GAMER-2}}: A {{GPU-accelerated}} Adaptive Mesh Refinement Code - Accuracy, Performance, and Scalability,} \mnras, 481, 4815, \dodoi{10.1093/mnras/sty2586}

% type= article
\bibitem[{E.~E. Schneider \& B.~E. Robertson(2015)Schneider \& Robertson}]{schneiderCHOLLA2015a}
Schneider, E.~E., \& Robertson, B.~E. 2015, \bibinfo{title}{{{CHOLLA}}: {{A NEW MASSIVELY PARALLEL HYDRODYNAMICS CODE FOR ASTROPHYSICAL SIMULATION}},} \apjs, 217, 24, \dodoi{10.1088/0067-0049/217/2/24}

% type= article
\bibitem[{P.~R. Shapiro \& H. Kang(1987)Shapiro \& Kang}]{shapiro1987hydrogen}
Shapiro, P.~R., \& Kang, H. 1987, \bibinfo{title}{Hydrogen molecules and the radiative cooling of pregalactic shocks,} \apj, 318, 32

% type= article
\bibitem[{J. Sirignano \& K. Spiliopoulos(2018)Sirignano \& Spiliopoulos}]{sirignanoDGM2018}
Sirignano, J., \& Spiliopoulos, K. 2018, \bibinfo{title}{{{DGM}}: {{A}} Deep Learning Algorithm for Solving Partial Differential Equations,} JCoPh, 375, 1339, \dodoi{10.1016/j.jcp.2018.08.029}

% type= article
\bibitem[{A. {Stacy} {et~al.}(2012){Stacy}, {Greif}, \& {Bromm}}]{Stacy:2012}
{Stacy}, A., {Greif}, T.~H., \& {Bromm}, V. 2012, \bibinfo{title}{{The first stars: mass growth under protostellar feedback},} \mnras, 422, 290, \dodoi{10.1111/j.1365-2966.2012.20605.x}

% type= article
\bibitem[{J.~M. Stone {et~al.}(2024)Stone, Mullen, Fielding, Grete, Guo, Kempski, Most, White, \& Wong}]{stoneAthenaK2024}
Stone, J.~M., Mullen, P.~D., Fielding, D., {et~al.} 2024, \bibinfo{title}{{{AthenaK}}: {{A Performance-Portable Version}} of the {{Athena}}++ {{AMR Framework}},} \apjs~submitted, arXiv:2409.16053, \dodoi{10.48550/arXiv.2409.16053}

% type= article
\bibitem[{K. {Sugimura} {et~al.}(2020){Sugimura}, {Matsumoto}, {Hosokawa}, {Hirano}, \& {Omukai}}]{Sugimura:2020}
{Sugimura}, K., {Matsumoto}, T., {Hosokawa}, T., {Hirano}, S., \& {Omukai}, K. 2020, \bibinfo{title}{{The Birth of a Massive First-star Binary},} \apjl, 892, L14, \dodoi{10.3847/2041-8213/ab7d37}

% type= article
\bibitem[{K. {Sugimura} {et~al.}(2023){Sugimura}, {Matsumoto}, {Hosokawa}, {Hirano}, \& {Omukai}}]{Sugimura:2023}
{Sugimura}, K., {Matsumoto}, T., {Hosokawa}, T., {Hirano}, S., \& {Omukai}, K. 2023, \bibinfo{title}{{Formation of Massive and Wide First-star Binaries in Radiation Hydrodynamic Simulations},} \apj, 959, 17, \dodoi{10.3847/1538-4357/ad02fc}

% type= article
\bibitem[{K. Sugimura {et~al.}(2014)Sugimura, Omukai, \& Inoue}]{sugimuraCritical2014}
Sugimura, K., Omukai, K., \& Inoue, A.~K. 2014, \bibinfo{title}{The Critical Radiation Intensity for Direct Collapse Black Hole Formation: Dependence on the Radiation Spectral Shape,} \mnras, 445, 544, \dodoi{10.1093/mnras/stu1778}

% type= article
\bibitem[{I. Sulzer \& T. Buck(2023)Sulzer \& Buck}]{sulzerSpeeding2023}
Sulzer, I., \& Buck, T. 2023, \bibinfo{title}{Speeding up Astrochemical Reaction Networks with Autoencoders and Neural {{ODEs}},} arXiv e-prints, arXiv:2312.06015, \dodoi{10.48550/arXiv.2312.06015}

% type= article
\bibitem[{H. {Susa}(2013){Susa}}]{Susa:2013}
{Susa}, H. 2013, \bibinfo{title}{{The Mass of the First Stars},} \apj, 773, 185, \dodoi{10.1088/0004-637X/773/2/185}

% type= article
\bibitem[{K.~E.~I. {Tanaka} \& K. {Omukai}(2014){Tanaka} \& {Omukai}}]{Tanaka:2014}
{Tanaka}, K. E.~I., \& {Omukai}, K. 2014, \bibinfo{title}{{Gravitational instability in protostellar discs at low metallicities},} \mnras, 439, 1884, \dodoi{10.1093/mnras/stu069}

% type= article
\bibitem[{R. Tibshirani(1996)Tibshirani}]{tibshiraniRegression1996a}
Tibshirani, R. 1996, \bibinfo{title}{Regression {{Shrinkage}} and {{Selection Via}} the {{Lasso}},} Journal of the Royal Statistical Society Series B: Statistical Methodology, 58, 267, \dodoi{10.1111/j.2517-6161.1996.tb02080.x}

% type= article
\bibitem[{P. van~de Bor {et~al.}(2025)van~de Bor, Brennan, Regan, \& Mackey}]{borBridging2025}
van~de Bor, P., Brennan, J., Regan, J.~A., \& Mackey, J. 2025, \bibinfo{title}{Bridging {Machine} {Learning} and {Cosmological} {Simulations}: Using {Neural} {Operators} to emulate {Chemical} {Evolution},} OJAp, 8, \dodoi{10.33232/001c.142225}

% type= article
\bibitem[{V. Wakelam {et~al.}(2012)Wakelam, Herbst, Loison, Smith, Chandrasekaran, Pavone, Adams, {Bacchus-Montabonel}, Bergeat, B{\'e}roff, Bierbaum, Chabot, Dalgarno, Van~Dishoeck, Faure, Geppert, Gerlich, Galli, H{\'e}brard, Hersant, Hickson, Honvault, Klippenstein, Le~Picard, Nyman, Pernot, Schlemmer, Selsis, Sims, Talbi, Tennyson, Troe, Wester, \& Wiesenfeld}]{wakelamKINETIC2012}
Wakelam, V., Herbst, E., Loison, J.-C., {et~al.} 2012, \bibinfo{title}{A {{KINETIC DATABASE FOR ASTROCHEMISTRY}} ({{KIDA}}),} The Astrophysical Journal Supplement Series, 199, 21, \dodoi{10.1088/0067-0049/199/1/21}

% type= article
\bibitem[{B.~D. {Wibking} \& M.~R. {Krumholz}(2022){Wibking} \& {Krumholz}}]{Wibking:2022}
{Wibking}, B.~D., \& {Krumholz}, M.~R. 2022, \bibinfo{title}{{QUOKKA: a code for two-moment AMR radiation hydrodynamics on GPUs},} \mnras, 512, 1430, \dodoi{10.1093/mnras/stac439}

% type= article
\bibitem[{N. Yoshida {et~al.}(2008)Yoshida, Omukai, \& Hernquist}]{yoshidaProtostar2008}
Yoshida, N., Omukai, K., \& Hernquist, L. 2008, \bibinfo{title}{Protostar {{Formation}} in the {{Early Universe}},} Science, 321, 669, \dodoi{10.1126/science.1160259}

\end{thebibliography}
\end{document}